\documentstyle[12pt,a4wide,amssymb,amsbsy,epsfig]{article}
\catcode`\@=11
\@addtoreset{equation}{section}

\begin{document}
\flushbottom
\pagestyle{empty}
\setcounter{page}{0}
\begin{flushright}
DFTT 63/94
\\
hep-ph/9411373
\end{flushright}
\vspace*{1cm}
\begin{center}
\Large \bf
A Model Independent Approach
\\
\vspace{0.3cm}
to Solar Neutrino
Experiments$^{\mbox{\footnotesize\mediumseries$\bigstar$}}$
\vspace*{1cm}
\\
\large \mediumseries
S.M. Bilenky$^{\mbox{\footnotesize\mediumseries(a,b)$\dagger$}}$
and
C. Giunti$^{\mbox{\footnotesize\mediumseries(a)$\ddagger$}}$
\\
\vspace{0.5cm}
\large
$^{\mbox{\footnotesize\mediumseries(a)}}$
INFN, Sezione di Torino,
and
Dipartimento di Fisica Teorica,
\\
Universit\`a di Torino,
Via P. Giuria 1, I--10125 Torino, Italy
\\
$^{\mbox{\footnotesize\mediumseries(b)}}$
Joint Institute for Nuclear Research, Dubna, Russia
\\
\vspace*{1cm}
Abstract
\\
\vspace{0.5cm}
\normalsize
\begin{minipage}[t]{0.9\textwidth}
In the first part of this report
we present
the results of a model independent analysis
of the existing solar neutrino data.
We obtained forbidden regions
in the plane of the parameters
$ \Delta m^2 $ and $ \sin^2 2 \theta $
in two cases:
A) Without any restrictions on the values
of the solar neutrino fluxes from different reactions;
B) With some restrictions
that take into account the predictions of all the existing solar models.
We show that the existing solar neutrino data
allow to exclude rather large regions
in the plane of the parameters
$ \Delta m^2 $ and $ \sin^2 2 \theta $
(especially in case B).
In the second part of this report
we present
a general method
for the analysis of solar neutrino data
that can be applied to future solar neutrino experiments
(SNO, Super-Kamiokande, Icarus)
in which high energy $^8\mathrm{B}$ neutrinos
will be detected.
We show that these experiments will allow:
1) To reveal in a model independent way
the presence of sterile neutrinos
in the flux of solar neutrinos on the earth
and to obtain lower bounds
for the probability of transition
of $\nu_e$'s into sterile states;
2) To obtain directly from the experimental data
the initial $^8\mathrm{B}$ $\nu_e$ flux
and the probability of $\nu_e$'s to survive
(if there are no transitions of $\nu_e$'s
into sterile states).
\end{minipage}
\end{center}
\vfill
\vspace*{1cm}
\begin{flushleft} \footnotesize
$\mbox{}^{\mbox{\footnotesize\mediumseries$\bigstar$}}$
Published in the Proceedings of
{\it The first Artic Workshop
on Future Physics and Accelerators},
Saariselka, Finland,
August 21--27, 1994,
p.~170--192.
\\
$\mbox{}^{\mbox{\footnotesize\mediumseries$\dagger$}}$
E-mail address: \texttt{bilenky@to.infn.it}
\\
$\mbox{}^{\mbox{\footnotesize\mediumseries$\ddagger$}}$
E-mail address: \texttt{giunti@to.infn.it}
\end{flushleft}

\newpage
\pagestyle{plain}

\section{Introduction}
\label{S1}

Solar neutrino experiments
are very important
for the investigation
of neutrino mixing,
as well as
for the investigation of the sun.
These experiments
are sensitive to very small values of the
neutrino mass difference squared
$ \Delta m^2 \equiv m_2^2 - m_1^2 $
(down to $ \Delta m^2 \simeq 10^{-10} \, \mathrm{eV}^2 $)
and
to a wide region
of mixing angles $ \theta $,
including very small
$ \theta $'s.
On the other side,
solar neutrino experiments
allow us to detect neutrinos from different reactions
of the thermonuclear solar cycles,
including neutrinos from
$^8\mathrm{B}$
decay,
whose flux is about $10^{-4}$
of the total flux.
The problem is that we cannot
determine from the existing solar neutrino experiments
[\ref{L:HOMESTAKE}--\ref{L:KAMIOKANDE}]
separately
the values of neutrino masses and mixing angles
{\em and}
the values of the neutrino fluxes.
Usually,
to obtain information
about the values of
$ \Delta m^2 $ and $ \sin^2 2 \theta $,
it is assumed that
the values of the neutrino fluxes are given by
the Standard Solar Model (SSM)
[\ref{L:BAHCALL}--\ref{L:NICE}].
It is well known,
however,
that the neutrino fluxes
calculated in the framework of the SSM
are subject to many sources of uncertainties,
mainly due to a poor knowledge
of some input parameters
(especially nuclear cross sections
and opacity).

In this report we will present:

\begin{enumerate}

\item
The results of
a solar model independent analysis
of the existing solar neutrino data~\cite{B:BG94B}.

\item
A model independent approach
to future solar neutrino experiments
in which solar neutrinos
will be detected
through the observation
of CC, NC and neutrino-electron
elastic scattering (ES)
processes
\cite{B:BG93,B:BG94A}.

\end{enumerate}

We will show that future solar neutrino experiments
(SNO~\cite{B:SNO},
Super-Kamiokande~\cite{B:SK},
ICARUS~\cite{B:ICARUS}),
in which high energy
$^8\mathrm{B}$ neutrinos
will be detected,
will allow to answer
in a model independent way
the question whether
there are transitions of solar $\nu_e$'s
into sterile states.
If there are only active neutrinos
$\nu_{e}$,
$\nu_{\mu}$,
$\nu_{\tau}$,
in the flux of solar neutrinos on the earth,
future solar neutrino experiments
will allow to determine
directly from the experimental data
the initial flux of
$^8\mathrm{B}$ $\nu_e$'s
and the probability
of $\nu_e$'s
to survive
as a function of neutrino energy.

There exist at present
data of three radiochemical
solar neutrino experiments
(Homestake
\cite{B:HOMESTAKE},
GALLEX
\cite{B:GALLEX}
and SAGE
\cite{B:SAGE})
and the water Cherenkov
direct counting Kamiokande experiment
\cite{B:KAMIOKANDE}.
In the Homestake experiment
solar neutrinos are detected
through the observation
of the Pontecorvo-Davis reaction
$ \nu_e + \mbox{}^{37}\mathrm{Cl}
\to
e^{-} + \mbox{}^{37}\mathrm{Ar} $,
whose threshold is
$ 0.81 \, \mathrm{MeV} $.
Thus $pp$ neutrinos,
which compose the main part of the solar neutrino flux,
are not detected in this experiment
and the main contribution
to the Homestake event rate
comes from
$^8\mathrm{B}$
and
$^7\mathrm{Be}$
(according to BP
\cite{B:BAHCALL}
78\% and 15\%,
respectively).
In the GALLEX and SAGE experiments
solar neutrinos are detected
through the observation
of the reaction
$ \nu_e + \mbox{}^{71}\mathrm{Ga}
\to
e^{-} + \mbox{}^{71}\mathrm{Ge} $,
whose threshold is
0.23 MeV.
The main contributions
to the GALLEX and SAGE
event rates come
from $pp$,
$^7\mathrm{Be}$
and
$^8\mathrm{B}$
(according to BP
54\%, 27\% and 8\%,
respectively).
In the Kamiokande experiment
solar neutrinos are detected
through the observation
of the process
$ \nu e \to \nu e $.
The electron energy threshold
in this experiment
is about 7 MeV.
Thus only
$^8\mathrm{B}$
neutrinos
are detected in the Kamiokande experiment.
In Table~\ref{T:ESND}
the results of all four solar neutrino experiments
are given.
In the last three columns
of Table~\ref{T:ESND}
the values of the event rates predicted by
Bahcall and Pinsonneault (BP)
\cite{B:BAHCALL},
Turck-Chi\`eze and Lopes (TL)
\cite{B:SACLAY}
and
Castellani, Degl'Innocenti and Fiorentini (CDF)
\cite{B:CDF}
are given.
As it is seen from Table~\ref{T:ESND},
the event rates measured in all four solar neutrino experiments
are significantly less than
the event rates
predicted by the existing
Standard Solar Models.

\begin{table}[t]
\begin{tabular*}{\textwidth}
{c@{\extracolsep{\fill}}
 c@{\extracolsep{\fill}}
 c@{\extracolsep{\fill}}
 c@{\extracolsep{\fill}}
 c}
\hline
\hline
\\
Experiment
&
Event Rate (SNU)
&
\multicolumn{3}{c}{SSM Predictions (SNU)}
\\
\\
&
&
BP
&
TL
&
CDF
\\
\\
\hline
\\
Homestake
&
$
N_{\mathrm{HOM}}^{\mathrm{exp}}
=
2.32 \pm 0.23
$
&
$ 8.0 \pm 1.0 $
&
$ 6.4 \pm 1.4 $
&
$ 7.8 $
\\
\\
GALLEX
&
$
N_{\mathrm{GAL}}^{\mathrm{exp}}
=
79 \pm 10 \pm 6
$
&
$ 131.5^{+7}_{-6} $
&
$ 123 \pm 7 $
&
$ 131 $
\\
\\
SAGE
&
$
N_{\mathrm{GAL}}^{\mathrm{exp}}
=
69 \pm 11 \pm 6
$
&
&
&
\\
\\
Kamiokande
&
$
N_{\mathrm{KAM}}^{\mathrm{exp}} / N_{\mathrm{KAM}}^{\mathrm{BP}}
=
0.51 \pm 0.04 \pm 0.06
$
&
$ 1 \pm 0.14 $
&
$ 0.8 \pm 0.2 $
&
$ 0.98 $
\\
\\
\hline
\hline
\end{tabular*}
\protect\caption{\small
Data of solar neutrino experiments
and rates predicted by
BP~\protect\cite{B:BAHCALL},
TL~\protect\cite{B:SACLAY}
and
CDF~\protect\cite{B:CDF}.
The Kamiokande result
is presented
as the ratio of the observed rate
to the rate predicted by BP.}
\label{T:ESND}
\end{table}

Pontecorvo neutrino mixing
(see, for example, Refs.\cite{B:BILENKY,B:CWKIM})
is apparently
the most natural explanation
of the possible
``deficit'' of solar neutrinos.
It was shown in
Refs.[\ref{L:GALLEX92B}--\ref{L:FLMMV94}]
that all existing solar neutrino data
can be described by the resonant MSW mechanism
\cite{B:MSW}
in the simplest case of mixing between
two neutrino types\footnote{
The data can be described also
by vacuum oscillations~\cite{B:KP}
with
a fine-tuning between $\Delta m^2$
and the sun-earth distance.
For the parameters
$ \Delta m^2 $ and $ \sin^2 2\theta $
the following values were obtained:
$ \Delta m^2 \simeq 8 \times 10^{-11} \, \mathrm{eV}^2 $
and
$ \sin^2 2\theta \simeq 0.8 $.
}.
The following
two MSW solutions of the solar neutrino problem
were found
in the case of $\nu_e$--$\nu_{\mu(\tau)}$
mixing:

\begin{enumerate}

\item
A small mixing angle solution:
$ \Delta m^2 \simeq 5 \times 10^{-6} \, \mathrm{eV}^2 $
and
$ \sin^2 2\theta \simeq 8 \times 10^{-3} $

\item
A large mixing angle solution:
$ \Delta m^2 \simeq 10^{-5} \, \mathrm{eV}^2 $
and
$ \sin^2 2\theta \simeq 0.8 $.

\end{enumerate}

Thus,
some indications in favor
of neutrino mixing
follow from the existing solar neutrino data.
We would like to emphasize,
however,
that these indications are model dependent:
the analysis of the data is based on the assumption
that the values of the neutrino fluxes
from the different sources
are given by the
Standard Solar Models.

\section{A model independent analysis of solar neutrino data}

In this section we present the results of a model independent
analysis
\cite{B:BG94B}
of the data of the existing solar neutrino experiments.
We will consider the simplest case
of mixing between two types of active neutrinos
($\nu_{e}$--$\nu_{\mu(\tau)}$)
and assume that the MSW mechanism takes place.
Our analysis is based on the fact that
the shapes of the spectra
of neutrinos from the reactions
of the thermonuclear cycle are known.
These spectra are determined by the interactions
responsible for the reactions
and,
as it was shown in Ref.\cite{B:BAHCALL91},
are negligibly affected
by the conditions in the interior of the sun.
Thus,
the event rates of the solar neutrino experiments
are determined
by the values of the parameters
$ \Delta m^2 $ and $ \sin^2 2\theta $
and by the values of the total neutrino fluxes
(mainly $pp$, $^7\mathrm{Be}$ and $^8\mathrm{B}$).
We will consider
the total neutrino fluxes
as unknown parameters.
{}From the existing solar neutrino data
we cannot determine {\em allowed} regions
of the values of the parameters
$ \Delta m^2 $ and $ \sin^2 2\theta $
{\em and} of the neutrino fluxes.
Instead,
we will obtain the regions of values
of the parameters
$ \Delta m^2 $ and $ \sin^2 2\theta $
that are {\em forbidden}
for all possible values
of the initial neutrino fluxes
(or for the initial neutrino fluxes
constrained within wide limits
chosen in such a way to include the predictions
of all the existing Standard Solar Models).
We will show that the existing solar neutrino data
allow to exclude rather large regions
of the values of the parameters
$ \Delta m^2 $ and $ \sin^2 2\theta $.

Let us write the initial
spectrum of $\nu_{e}$
from the source $r$
($r$ $=$
$pp$, $pep$, $^7\mathrm{Be}$, $^8\mathrm{B}$, $\mathrm{He}p$,
$^{13}\mathrm{N}$, $^{15}\mathrm{O}$, $^{17}\mathrm{F}$)
in the form
\begin{equation}
\phi_{\nu_{e}}^{r}(E)
=
X^{r}(E)
\,
\Phi^{r}
\;,
\label{E1011}
\end{equation}
where
$E$ is the neutrino energy,
$ X^{r}(E) $
is a known function
normalized by the condition
$ \displaystyle \int X^{r}(E) \, {\mathrm{d}} E = 1 $
(see Ref.\cite{B:BAHCALL})
and
$ \Phi^{r} $
is the total initial neutrino flux
from the source $r$.

The integral event rate in any experiment $a$
($a$ = HOM (Homestake), KAM (Kamiokande), GAL (GALLEX+SAGE)\footnote{
In our calculations we use the combined
GALLEX--SAGE data: $ 74 \pm 9 $ SNU.
})
is given by the expression
\begin{equation}
N_{a}
=
\sum_{r}
Y_{a}^{r}
\,
\Phi^{r}
\;.
\label{E1001}
\end{equation}
In the case of radiochemical experiments
only solar $\nu_{e}$ are detected.
We have
\begin{equation}
Y_{a}^{r}
=
\int_{E_{\mathrm{th}}^{a}}
\sigma_{a}(E)
\,
X^{r}(E)
\,
\mathrm{P}_{\nu_{e}\to\nu_{e}}(E)
\,
\mathrm{d} E
\;,
\label{E1004}
\end{equation}
with
$a=\mathrm{HOM},\mathrm{GAL}$.
Here
$ \sigma_{a}(E) $
is
the cross section
of the reaction
$ \nu_e + \mbox{}^{37}\mathrm{Cl}
\to
e^{-} + \mbox{}^{37}\mathrm{Ar} $
in the case of the Homestake experiment
and
the cross section of the reaction
$ \nu_e + \mbox{}^{71}\mathrm{Ga}
\to
e^{-} + \mbox{}^{71}\mathrm{Ge} $,
in the case of the GALLEX and SAGE experiments,
$ \mathrm{P}_{\nu_{e}\to\nu_{e}}(E) $
is the probability of $\nu_{e}$ to survive
and
$ E_{\mathrm{th}}^{a} $
is the threshold energy.
For the calculation of
the $\nu_{e}$ survival probability
we used the formula given in Ref.\cite{B:PEEMSW},
which is valid for an exponentially decreasing
electron density.

In the the Kamiokande experiment
$\nu_{e}$ as well as $\nu_{\mu}$ (and/or $\nu_{\tau}$)
are detected.
We have
\begin{eqnarray} \nopagebreak
&&
Y_{\mathrm{KAM}}^{r}
=
\int_{E_{\mathrm{th}}^{\mathrm{ES}}}
\sigma_{\nu_{e}e}(E)
\,
X^{r}(E)
\,
\mathrm{P}_{\nu_{e}\to\nu_{e}}(E)
\,
\mathrm{d} E
\nonumber
\\
&&
\phantom{ Y_{\mathrm{KAM}}^{r} }
+
\int_{E_{\mathrm{th}}^{\mathrm{ES}}}
\sigma_{\nu_{\mu}e}(E)
\,
X^{r}(E)
\,
\sum_{\ell=\mu,\tau} \mathrm{P}_{\nu_{e}\to\nu_{\ell}}(E)
\,
\mathrm{d} E
\;,
\label{E1003}
\end{eqnarray}
where
$ \sigma_{\nu_\ell e}(E) $
is the cross section of the process
$ \nu_\ell \, e \to \nu_\ell \, e $
($\ell=e,\mu$),
$ E_{\mathrm{th}}^{\mathrm{ES}} $
is the recoil electron energy threshold
and
$ \mathrm{P}_{\nu_{e}\to\nu_{\ell}}(E) $
is the probability of the transition
$ \nu_{e} \to \nu_{\ell} $
($ \ell = e , \mu $).
In our calculation we took into account
the efficiency and the energy resolution of the Kamiokande detector
\cite{B:KAMIOKANDE}.
The fluxes of neutrinos produced
in the thermonuclear $pp$ and CNO
cycles must satisfy the following relation:
\begin{equation}
N_{\mathrm{LUM}}
=
\sum_{r}
Y_{\mathrm{LUM}}^{r}
\,
\Phi^{r}
\;.
\label{E1012}
\end{equation}
Here
$ \displaystyle
N_{\mathrm{LUM}}
=
L_{\odot}
/
4 \pi \, d^2
=
( 8.491 \pm 0.018 ) \times 10^{11}
\, \mathrm{MeV} \, \mathrm{cm}^{-2} \, \mathrm{sec}^{-1}
$,
where
$ d = 1 \, \mathrm{AU} = 1.496 \times 10^{13} \, \mathrm{cm} $
is the average sun-earth distance,
$
L_{\odot}
=
( 3.826 \pm 0.008 ) \times 10^{33}
\, \mathrm{erg}
\, \mathrm{sec}^{-1}
$
\cite{B:PDG92},
is the luminosity of the sun
and
$ \displaystyle
Y_{\mathrm{LUM}}^{r}
=
Q / 2 - \left\langle E \right\rangle^{r}
$,
where
$ Q = 4 \, m_{p} + 2 \, m_{e} - m_{^4\mathrm{He}} = 26.73 \, \mathrm{MeV} $
and
$ \left\langle E \right\rangle^{r} $
is the average energy of neutrinos from the source $r$.
The values of
$ \left\langle E \right\rangle^{r} $
and
$ Y_{\mathrm{LUM}}^{r} $
are given in Table~\ref{T:FLUXES}.

\begin{table}[t]
\begin{tabular*}{\textwidth}
{c@{\extracolsep{\fill}}
 c@{\extracolsep{\fill}}
 c@{\extracolsep{\fill}}
 c@{\extracolsep{\fill}}
 c@{\extracolsep{\fill}}
 c}
\hline
\hline
Source
&
$ \displaystyle
\begin{array}{c} \displaystyle
\left\langle E \right\rangle
\\ \displaystyle
(\mathrm{MeV})
\end{array}
$
&
$ \displaystyle
\begin{array}{c} \displaystyle
Y_{\mathrm{LUM}}
\\ \displaystyle
(\mathrm{MeV})
\end{array}
$
&
$ \displaystyle
\begin{array}{c} \displaystyle
\Phi(\mathrm{BP})
\\ \displaystyle
(\mathrm{cm}^{-2} \mathrm{sec}^{-1})
\end{array}
$
&
$ \xi_{\mathrm{min}} $
&
$ \xi_{\mathrm{max}} $
\\
\hline
$pp$
&
0.265
&
13.10
&
$ ( 6.00 \pm 0.004 ) \times 10^{10} $
&
0.93
&
1.07
\\
$pep$
&
1.442
&
11.92
&
$ ( 1.43 \pm 0.02 ) \times 10^{8} $
&
0.61
&
1.29
\\
$^7\mathrm{Be}$
&
0.813
&
12.55
&
$ ( 4.89 \pm 0.29 ) \times 10^{9} $
&
0.46
&
1.40
\\
$^8\mathrm{B}$
&
6.710
&
6.66
&
$ ( 5.69 \pm 0.82 ) \times 10^{6} $
&
0
&
1.43
\\
$\mathrm{He}p$
&
9.625
&
3.74
&
$ 1.23 \times 10^{3} $
&
0.90
&
1.13
\\
$^{13}\mathrm{N}$
&
0.7067
&
12.66
&
$ ( 4.92 \pm 0.84 ) \times 10^{8} $
&
0
&
1.51
\\
$^{15}\mathrm{O}$
&
0.9965
&
12.37
&
$ ( 4.26 \pm 0.82 ) \times 10^{8} $
&
0
&
1.58
\\
$^{17}\mathrm{F}$
&
0.9994
&
12.37
&
$ ( 5.39 \pm 0.86 ) \times 10^{6} $
&
0
&
1.48
\\
\hline
\hline
\end{tabular*}
\protect\caption{\small
Solar neutrino fluxes (with $1\sigma$ errors) predicted by BP;
$ \left\langle E \right\rangle $
is the average neutrino energy,
$ Y_{\mathrm{LUM}} = Q/2 - \left\langle E \right\rangle$,
where
$ Q = 26.73 \, \mathrm{MeV} $,
and
$ \xi_{\mathrm{min}} $
and
$ \xi_{\mathrm{max}} $
determine the limits for the values of the total neutrino fluxes
in case B.}
\label{T:FLUXES}
\end{table}

\begin{figure}[t]
\begin{minipage}[t]{0.49\linewidth}
\begin{center}
\mbox{\epsfig{file=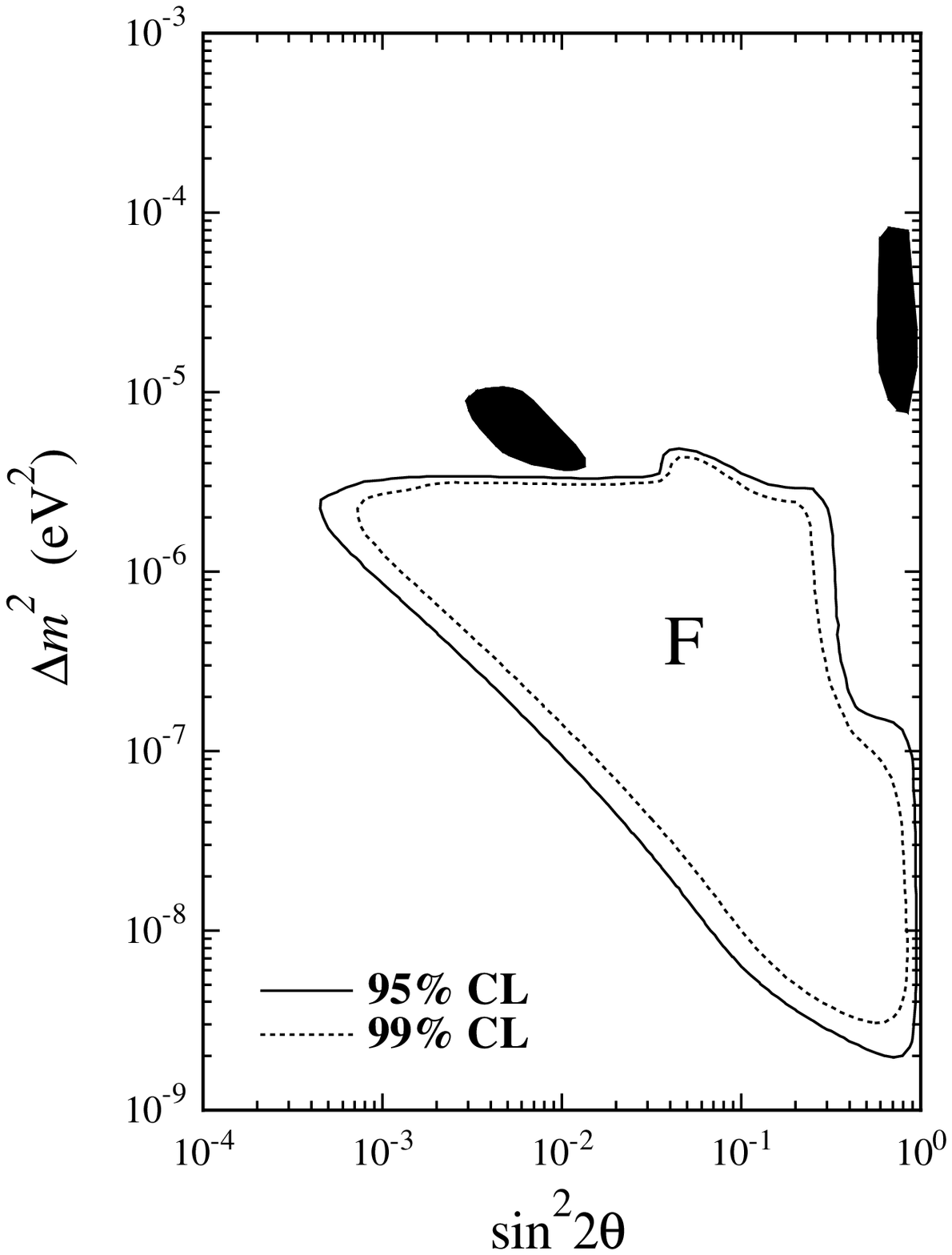,width=\linewidth}}
\end{center}
\protect\caption{\small
Excluded regions in the
$ \sin^2 2\theta $--$ \Delta m^2 $
plane for MSW transitions
due to
$\nu_{e}$--$\nu_{\mu(\tau)}$
mixing in case A.
The region F
is excluded at 95\% CL
within the solid line
and at 95\% CL
within the dotted line.
The allowed regions
found with the BP neutrino fluxes are also shown (shaded areas).
}
\label{F01}
\end{minipage}
\hfill
\begin{minipage}[t]{0.49\linewidth}
\begin{center}
\mbox{\epsfig{file=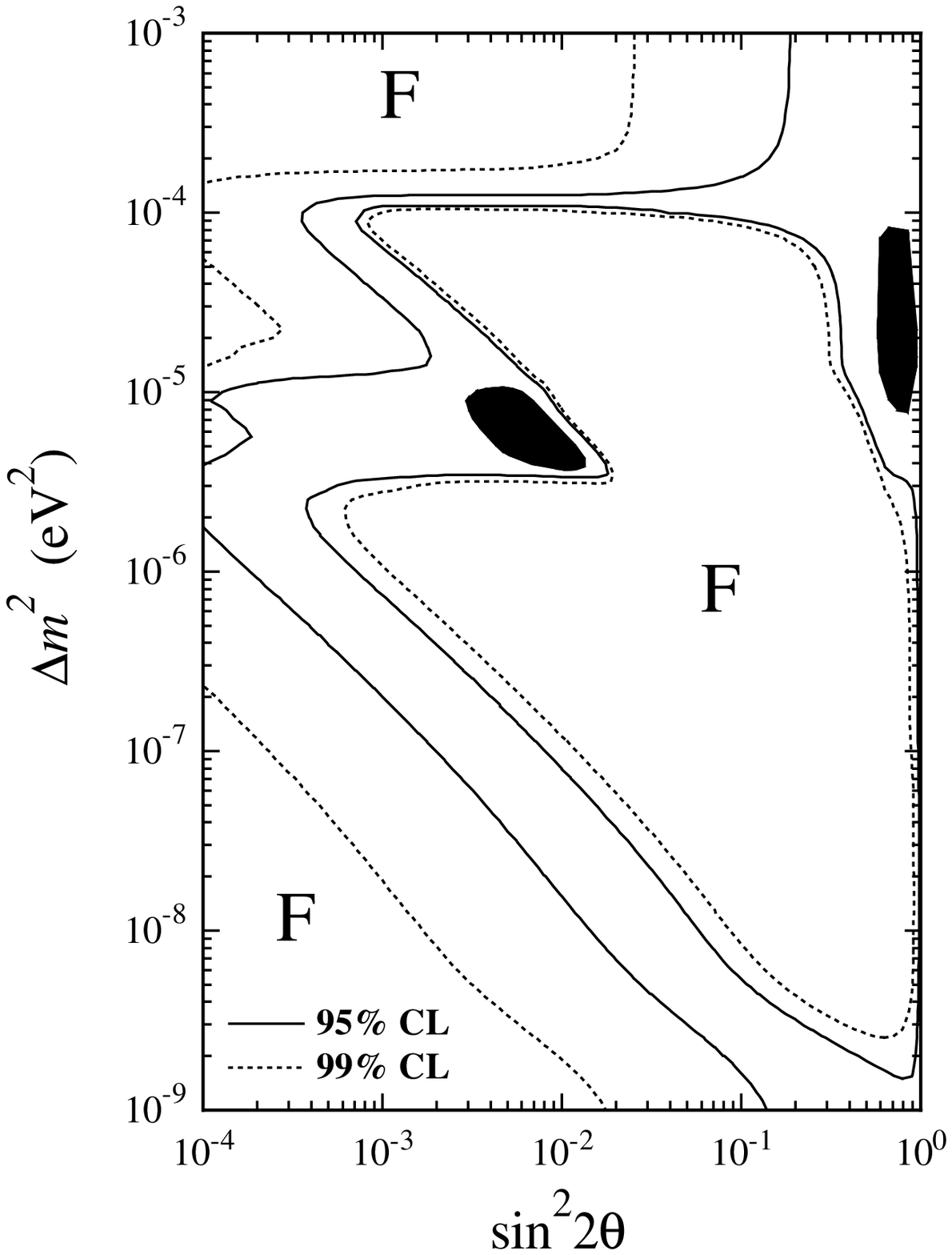,width=\linewidth}}
\end{center}
\protect\caption{\small
Excluded regions in the
$ \sin^2 2\theta $--$ \Delta m^2 $
plane for MSW transitions
due to
$\nu_{e}$--$\nu_{\mu(\tau)}$
mixing in case B.
The regions F
are excluded at 95\% CL
within the corresponding solid line
and at 95\% CL
within the corresponding dotted line.
The allowed regions
found with the BP neutrino fluxes are also shown (shaded areas).
}
\label{F02}
\end{minipage}
\end{figure}

Our procedure for the analysis of the solar neutrino data
is the following.
At fixed values of the parameters
$ \Delta m^2 $ and $ \sin^2 2\theta $
we calculate the $\chi^2$
for all possible values of the neutrino fluxes.
For each value
of the parameters
$ \Delta m^2 $ and $ \sin^2 2\theta $
and of the neutrino fluxes
we estimate the ``goodness-of-fit''
by calculating the confidence level (CL)
corresponding to the calculated $\chi^2$.
Since we do not determine any parameter,
the number of degrees of freedom
of the $\chi^2$ distribution
is equal to the number of data points
(i.e. four: three neutrino rates and the solar luminosity constraint).
If all the confidence levels
found for a given value of
$ \Delta m^2 $, $ \sin^2 2\theta $
and all possible values of the neutrino fluxes
are smaller than $\alpha$
(we choose $\alpha$ = 0.1, 0.05, 0.01),
then the corresponding point in the
$ \Delta m^2 $--$ \sin^2 2\theta $
plane is excluded at
$ 100 ( 1 - \alpha ) $\% CL.
In this way we obtain the exclusion plots
presented in Figs.\ref{F01} and \ref{F02}.
Let us notice that for the purpose of determination
of the excluded regions
in the parameter space
our approach is the most conservative:
any decrease of the number of degrees of freedom
would increase the excluded regions.

For the exclusion plot presented in
Fig.\ref{F01}
the only requirement was that all the total neutrino fluxes
are positive.
Let us call this case A.
However,
it is interesting and instructive to investigate
how the forbidden regions in the
$ \Delta m^2 $--$ \sin^2 2\theta $ plane
change if some limits on the allowed
values of the neutrino fluxes are imposed.
Thus we also considered the following case B:
the different solar neutrino fluxes
are constrained in the interval
$
\xi^{r}_{\mathrm{min}} \, \Phi^{r}(\mathrm{BP})
\le
\Phi^{r}
\le
\xi^{r}_{\mathrm{max}} \, \Phi^{r}(\mathrm{BP})
$,
where
$ \Phi^{r}(\mathrm{BP}) $
are the BP values of the neutrino fluxes
and the factors
$ \xi^{r}_{\mathrm{min}} $
and
$ \xi^{r}_{\mathrm{max}} $
are chosen
in such a way to include the predictions
of the existing solar models
[\ref{L:BAHCALL}--\ref{L:NICE}].
The values of these factors are given in Table~\ref{T:FLUXES}.
We determined the minimum (maximum) values
for the $pp$, $pep$, $^7\mathrm{Be}$ and $\mathrm{He}p$ fluxes
by subtracting (adding) 3 times the range of solar model predictions
to the minimum (maximum) predicted flux
(notices that this range is larger
than the $1\sigma$
error given by BP).
Since it has been recently suggested
\cite{B:S17}
that
the value of the astrophysical factor
$S_{17}(0)$ could be significantly lower
than that used in SSM calculations,
we let the $^8\mathrm{B}$ flux
to be arbitrarily small.
Since the CNO fluxes
have large uncertainties,
we allow also them
to be arbitrarily small.
We determined the maximum values
of the $^8\mathrm{B}$ and CNO fluxes
by adding 3 times
the $1\sigma$
error of BP
to the BP average value.
Let us emphasize that the limits
on the allowed
values of the neutrino fluxes which we imposed in case B
are rather large.
The excluded regions
of the parameters
$ \Delta m^2 $ and $ \sin^2 2\theta $
in case B are presented in Fig.\ref{F02}.
As it can be seen from a comparison
of this figure with Fig.\ref{F01}
the excluded regions in case B
are much larger than in the case where no limitation is imposed
on the values of the neutrino fluxes.

\section{Are there sterile neutrinos
in the flux of solar neutrinos on the earth?}

The problem of existence of sterile neutrinos\footnote{
Sterile neutrinos
are quanta
of right-handed neutrino fields.
In the Dirac-Majorana mixing scheme,
these fields
are mixed with the left-handed fields
(see, for example, Refs.\cite{B:BILENKY,B:CWKIM}).
Sterile neutrinos
do not interact with matter
via standard CC and NC interactions.
}
is very important for the theory
beyond the Standard Model.
Neutrino masses and mixing
can be generated in the framework of the standard model
if right-handed neutrino fields (singlets)
together with left-handed doublets
enter in the Yukawa interaction.
In this case
the total lepton charge
$ L = L_{e} + L_{\mu} + L_{\tau} $
is conserved,
neutrinos with definite masses
are Dirac particles
and
transitions only between flavor neutrinos
($ \nu_{\ell} \to \nu_{\ell'} $
with
$ \ell , \ell' = e , \mu , \tau $)
are allowed.
In the models
beyond the Standard Model
transitions of active flavor neutrinos into sterile neutrino states
become possible
(see, for example, Refs.\cite{B:BILENKY,B:CWKIM}).
So a discovery of such transitions
will be a discovery of new physics.

Here
we will show that future solar neutrino experiments could allow
to reveal the presence of sterile neutrinos
in the solar neutrino flux on the earth
independently on any assumption
about the values of the initial neutrino fluxes
\cite{B:BG93,B:BG94A}.

The solar neutrino experiments of the next generation
(scheduled to start in 1996)
are
the SNO \cite{B:SNO}
and
the Super-Kamiokande \cite{B:SK}
experiments.
In the SNO experiment
solar neutrinos
will be detected through
the observation of {\em three}
different processes:

\begin{enumerate}

\item
The CC process
\begin{equation}
\nu_{e} + d \to e^{-} + p + p
\; ;
\label{E:CC}
\end{equation}

\item
The NC process
\begin{equation}
\nu + d \to \nu + p + n
\; ;
\label{E:NC}
\end{equation}

\item
The elastic scattering (ES) process
\begin{equation}
\nu + e^{-} \to \nu + e^{-}
\; .
\label{E:ES}
\end{equation}

\end{enumerate}

In the Super-Kamiokande (S-K)
experiment solar neutrinos
will be detected through the observation
of the process
(\ref{E:ES})
with an event rate
about 30 times larger than in the current Kamiokande experiment.
In both the SNO and S-K experiments,
due to the high energy thresholds
($\simeq 6 \, \mathrm{MeV}$
for the CC process,
$2.2 \, \mathrm{MeV}$
for the NC process
and
$\simeq 5 \, \mathrm{MeV}$
for the ES process),
only neutrinos coming from
$^8\mathrm{B}$ decay
will be detected.
The energy spectrum of the initial $^8\mathrm{B}$ $\nu_e$'s
is given by
\begin{equation}
\phi_{\nu_{e}}^{0}(E)
=
X(E)
\,
\Phi
\; .
\label{E500}
\end{equation}
The function $X(E)$ is the normalized
neutrino spectrum
from the decay
$ ^8\mathrm{B} \to \mbox{} ^8\mathrm{Be} + e^{+} + \nu_{e} $,
which is determined by the phase space factor
(small corrections due to forbidden transitions
where calculated in Ref.\cite{B:BH86}).
The factor
$\Phi$
in Eq.(\ref{E500})
is the total flux of initial $^8\mathrm{B}$ solar $\nu_{e}$'s.

Consider first the NC process (\ref{E:NC}).
Using
$\nu_{e}$--$\nu_{\mu}$--$\nu_{\tau}$
universality of NC
for the total NC event rate in the SNO experiment
$N^{\mathrm{NC}}$,
we have
\begin{equation}
N^{\mathrm{NC}}
=
\int_{E_{\mathrm{th}}^{\mathrm{NC}}}
\sigma_{{\nu}d}(E) \,
X(E)
\sum_{\ell=e,\mu,\tau} \mathrm{P}_{\nu_{e}\to\nu_{\ell}}(E) \,
\mathrm{d} E
\,
\Phi
\; ,
\label{E501}
\end{equation}
where
$ \sigma_{{\nu}d}(E) $
is the cross section for the process
$ \nu \, d \to \nu \, n \, p $,
$ E_{\mathrm{th}}^{\mathrm{NC}} $
is the threshold neutrino energy
and
$ \mathrm{P}_{\nu_{e}\to\nu_{\ell}}(E) $
is the probability
of transition of solar $\nu_e$'s
into $\nu_{\mathrm{\ell}}$
($ \ell = e , \mu , \tau $).
It is useful to introduce the average total
probability
of transitions of $\nu_{e}$
into other active neutrinos:
\begin{equation}
\left\langle
\sum_{\ell=e,\mu,\tau} \mathrm{P}_{\nu_{e}\to\nu_{\ell}}
\right\rangle_{\mathrm{NC}}
\equiv
{\displaystyle
1
\over\displaystyle
X_{{\nu}d}
}
\int_{E_{\mathrm{th}}^{\mathrm{NC}}}
\sigma_{{\nu}d}(E) \,
X(E)
\sum_{\ell=e,\mu,\tau} \mathrm{P}_{\nu_{e}\to\nu_{\ell}}(E) \,
\mathrm{d} E
\; ,
\label{E502}
\end{equation}
where
\begin{equation}
X_{{\nu}d}
\equiv
\int_{E_{\mathrm{th}}^{\mathrm{NC}}}
\sigma_{{\nu}d}(E) \,
X(E) \,
{\mathrm{d}} E
\; .
\label{E599}
\end{equation}
Using the results of a recent calculation of the cross-section
$ \sigma_{{\nu}d}(E) $
\cite{B:KN93}
we obtained
$ X_{{\nu}d} = 4.72 \times 10^{-43} \,\mathrm{cm}^2 $.
{}From Eqs.(\ref{E501}) and (\ref{E502})
we get
\begin{equation}
\left\langle
\sum_{\ell=e,\mu,\tau} \mathrm{P}_{\nu_{e}\to\nu_{\ell}}
\right\rangle_{\mathrm{NC}}
=
{\displaystyle
N^{\mathrm{NC}}
\over\displaystyle
X_{{\nu}d}
\,
\Phi
}
\; .
\label{E503}
\end{equation}
In the general case of transitions
of $\nu_e$'s
into active as well as into sterile neutrinos
\begin{equation}
\sum_{\ell=e,\mu,\tau} \mathrm{P}_{\nu_{e}\to\nu_{\ell}}(E)
=
1
-
\mathrm{P}_{\nu_{e}\to\nu_{\mathrm{S}}}(E)
\; ,
\label{E504}
\end{equation}
where
$ \mathrm{P}_{\nu_{e}\to\nu_{\mathrm{S}}}(E) $
is the total
probability of transition of
$\nu_e$'s
into all possible sterile states\footnote{
We assume that neutrinos are stable particles.
For a discussion
of neutrino instability
see Ref.\cite{B:BMA}
and references therein.
}.
{}From Eqs.(\ref{E503}) and (\ref{E504})
it follows that
\begin{equation}
1
-
\left\langle
\mathrm{P}_{\nu_{e}\to\nu_{\mathrm{S}}}
\right\rangle_{\mathrm{NC}}
=
{\displaystyle
N^{\mathrm{NC}}
\over\displaystyle
X_{{\nu}d}
\,
\Phi
}
\; .
\label{E505}
\end{equation}
Thus,
from the measurement of the NC event rate
$ N^{\mathrm{NC}} $
it is impossible to reach any conclusions
about transitions of solar $\nu_e$'s
into sterile states without assumptions
about the value of the total flux
$ \Phi $.
However,
if
solar $^8\mathrm{B}$ neutrinos
are detected not only through the observation of NC
but also through the observation of the ES and CC processes
the problem of existence of sterile neutrinos
could be solved in a completely
model independent way.

Let us consider the ES process (\ref{E:ES}).
The total number of ES events is equal to
\begin{equation}
N^{\mathrm{ES}}
=
\int_{{E_{\mathrm{th}}^{\mathrm{ES}}}}
\sigma_{\nu_{e}e}(E)
\,
\mathrm{P}_{\nu_{e}\to\nu_{e}}(E)
\,
X(E)
\,
{\mathrm{d}} E
\,
\Phi
+
\int_{{E_{\mathrm{th}}^{\mathrm{ES}}}}
\sigma_{\nu_{\mu}e}(E)
\sum_{\ell=\mu,\tau}
\mathrm{P}_{\nu_{e}\to\nu_{\ell}}(E)
\,
X(E)
\,
{\mathrm{d}} E
\,
\Phi
\; .
\label{E506}
\end{equation}
{}From Eq.(\ref{E506}) we have
\begin{equation}
\Sigma^{\mathrm{ES}}
=
\int_{{E_{\mathrm{th}}^{\mathrm{ES}}}}
\sigma_{\nu_{\mu}e}(E)
\sum_{\ell=e,\mu,\tau}
\mathrm{P}_{\nu_{e}\to\nu_{\ell}}(E)
\,
X(E)
\,
{\mathrm{d}} E
\,
\Phi
\; .
\label{E507}
\end{equation}
Here
\begin{equation}
\Sigma^{\mathrm{ES}}
\equiv
N^{\mathrm{ES}}
-
\int_{{E_{\mathrm{th}}^{\mathrm{ES}}}}
\left(
\sigma_{\nu_{e}e}(E)
-
\sigma_{\nu_{\mu}e}(E)
\right)
\phi_{\nu_{e}}(E)
\,
{\mathrm{d}} E
\; ,
\label{E508}
\end{equation}
where
$
\phi_{\nu_{e}}(E)
=
\mathrm{P}_{\nu_{e}\to\nu_{e}}(E)
\,
X(E)
\,
\Phi
$
is the flux of $\nu_e$ on the earth.
The quantity
$ \Sigma^{\mathrm{ES}} $
can be obtained from the data
of the SNO and S-K experiments.
In fact,
$ N^{\mathrm{ES}} $
will be measured in both experiments.
In the SNO experiment
the spectrum of the electrons
in the CC process (\ref{E:CC})
will be measured and the spectrum
of solar $\nu_e$ on the earth,
$ \phi_{\nu_{e}}(E) $,
will be determined.
{}From Eq.(\ref{E507})
we obtain the relation
\begin{equation}
\left\langle
\sum_{\ell=e,\mu,\tau} \mathrm{P}_{\nu_{e}\to\nu_{\ell}}
\right\rangle_{\mathrm{ES}}
=
{\displaystyle
\Sigma^{\mathrm{ES}}
\over\displaystyle
X_{\nu_{\mu}e}
\,
\Phi
}
\; ,
\label{E509}
\end{equation}
which is similar in form to the relation (\ref{E503}).
The quantities
in the relation (\ref{E509})
are determined as follows:
\begin{equation}
X_{\nu_{\mu}e}
\equiv
\int_{E_{\mathrm{th}}^{\mathrm{ES}}}
\sigma_{\nu_{\mu}e}(E)
\,
X(E)
\,
{\mathrm{d}} E
\label{E510}
\end{equation}
and
\begin{equation}
\left\langle
\sum_{\ell=e,\mu,\tau} \mathrm{P}_{\nu_{e}\to\nu_{\ell}}
\right\rangle_{\mathrm{ES}}
\equiv
{\displaystyle
1
\over\displaystyle
X_{\nu_{\mu}e}
}
\int_{E_{\mathrm{th}}^{\mathrm{ES}}}
\sigma_{\nu_{\mu}e}(E)
\,
X(E)
\sum_{\ell=e,\mu,\tau} \mathrm{P}_{\nu_{e}\to\nu_{\ell}}(E)
\,
\mathrm{d} E
\; .
\label{E511}
\end{equation}
For
$ E_{\mathrm{th}}^{\mathrm{ES}} = 5.94 \, \mathrm{MeV} $
(which corresponds to a
kinetic energy threshold
$ \mathrm{T}_{\mathrm{th}} = 4.5 \, \mathrm{MeV} $
for the electrons in the CC process)
we have
$ \displaystyle
X_{\nu_{\mu}e}
=
2.08 \times 10^{-45} \, \mathrm{cm}^2
$.
{}From Eqs.(\ref{E503}) and (\ref{E509})
we have
\begin{equation}
{\displaystyle
1
-
\left\langle
\mathrm{P}_{\nu_{e}\to\nu_{\mathrm{S}}}
\right\rangle_{\mathrm{ES}}
\over\displaystyle
1
-
\left\langle
\mathrm{P}_{\nu_{e}\to\nu_{\mathrm{S}}}
\right\rangle_{\mathrm{NC}}
}
=
\mathrm{R}^{\mathrm{ES}}_{\mathrm{NC}}
\; ,
\label{E512}
\end{equation}
where
\begin{equation}
\mathrm{R}^{\mathrm{ES}}_{\mathrm{NC}}
\equiv
{\displaystyle
\Sigma^{\mathrm{ES}}
\,
X_{{\nu}d}
\over\displaystyle
X_{\nu_{\mu}e}
\,
N^{\mathrm{NC}}
}
\; .
\label{E513}
\end{equation}
Let us stress
that in the ratio
$ \mathrm{R}^{\mathrm{ES}}_{\mathrm{NC}} $
only measurable and known
quantities enter
(the flux
$ \Phi $
cancels in the ratio).
{}From Eq.(\ref{E512})
it is evident that
if
\begin{equation}
\mathrm{R}^{\mathrm{ES}}_{\mathrm{NC}}
\not=
1
\label{E514}
\end{equation}
it would mean that
there are transitions of
solar $\nu_e$'s into sterile states.

In the case
$ \mathrm{R}^{\mathrm{ES}}_{\mathrm{NC}} = 1 $
no conclusion about sterile neutrinos
can be reached.
In fact,
the ratio
$ \mathrm{R}^{\mathrm{ES}}_{\mathrm{NC}} $
is equal to one if
$ \displaystyle
\left\langle
\mathrm{P}_{\nu_{e}\to\nu_{\mathrm{S}}}
\right\rangle_{\mathrm{NC}}
=
\left\langle
\mathrm{P}_{\nu_{e}\to\nu_{\mathrm{S}}}
\right\rangle_{\mathrm{ES}}
$.
This relation is satisfied at
$ \mathrm{P}_{\nu_{e}\to\nu_{\mathrm{S}}}(E) = 0 $
as well as at
$ \mathrm{P}_{\nu_{e}\to\nu_{\mathrm{S}}}(E) = \mbox{const} \not= 0 $.
Thus,
if
$ \mathrm{R}^{\mathrm{ES}}_{\mathrm{NC}} \not= 1 $
it would mean
not only that sterile neutrinos exist,
but also that
the probability
of the transition of solar $\nu_e$'s
into sterile states depends on neutrino energy.

If the inequality (\ref{E514})
takes place,
it is possible to obtain lower bounds
for the average values of the probability
of transition
of $\nu_e$'s into sterile states.
In fact,
we have
\begin{equation}
\left\langle
\sum_{\ell=e,\mu,\tau} \mathrm{P}_{\nu_{e}\to\nu_{\ell}}
\right\rangle_{\mathrm{ES}}
\le
{\displaystyle
\left\langle
\sum_{\ell=e,\mu,\tau} \mathrm{P}_{\nu_{e}\to\nu_{\ell}}
\right\rangle_{\mathrm{ES}}
\over\displaystyle
\left\langle
\sum_{\ell=e,\mu,\tau} \mathrm{P}_{\nu_{e}\to\nu_{\ell}}
\right\rangle_{\mathrm{NC}}
}
=
\mathrm{R}^{\mathrm{ES}}_{\mathrm{NC}}
\; .
\label{E515}
\end{equation}
{}From this inequality
it follows that
\begin{equation}
\left\langle
\mathrm{P}_{\nu_{e}\to\nu_{\mathrm{S}}}
\right\rangle_{\mathrm{ES}}
\ge
1
-
\mathrm{R}^{\mathrm{ES}}_{\mathrm{NC}}
\; .
\label{E516}
\end{equation}
Thus,
if the inequality
\begin{equation}
\mathrm{R}^{\mathrm{ES}}_{\mathrm{NC}}
<
1
\label{E517}
\end{equation}
is satisfied,
from Eq.(\ref{E516})
we obtain a non-zero
lower bound for the average probability
$
\left\langle
\mathrm{P}_{\nu_{e}\to\nu_{\mathrm{S}}}
\right\rangle_{\mathrm{ES}}
$.

In the case of transitions of solar $\nu_e$'s
into sterile states
the initial $^8\mathrm{B}$ $\nu_e$ flux
cannot be determined from the experimental data.
However,
in this case
we can obtain from the data a lower bound
for this flux.
In fact,
from Eqs.(\ref{E509}), (\ref{E513}) and (\ref{E515})
we have
\begin{equation}
\Phi
\ge
{\displaystyle
\Sigma^{\mathrm{ES}}
\over\displaystyle
X_{\nu_{\mu}e}
\,
\mathrm{R}^{\mathrm{NC}}_{\mathrm{ES}}
}
=
{\displaystyle
N^{\mathrm{NC}}
\over\displaystyle
X_{{\nu}d}
}
\; .
\label{E518}
\end{equation}

Further,
we have
\begin{equation}
\left\langle
\sum_{\ell=e,\mu,\tau} \mathrm{P}_{\nu_{e}\to\nu_{\ell}}
\right\rangle_{\mathrm{NC}}
\le
{\displaystyle
\left\langle
\sum_{\ell=e,\mu,\tau} \mathrm{P}_{\nu_{e}\to\nu_{\ell}}
\right\rangle_{\mathrm{NC}}
\over\displaystyle
\left\langle
\sum_{\ell=e,\mu,\tau} \mathrm{P}_{\nu_{e}\to\nu_{\ell}}
\right\rangle_{\mathrm{ES}}
}
=
\left( \mathrm{R}^{\mathrm{ES}}_{\mathrm{NC}} \right)^{-1}
\; .
\label{E521}
\end{equation}
{}From Eq.(\ref{E521})
it follows that if the inequality
\begin{equation}
\mathrm{R}^{\mathrm{ES}}_{\mathrm{NC}}
>
1
\label{E522}
\end{equation}
is satisfied,
then for the average probability
$
\left\langle
\mathrm{P}_{\nu_{e}\to\nu_{\mathrm{S}}}
\right\rangle_{\mathrm{NC}}
$
and for the total flux
$ \Phi $
we have the following lower bounds:
\begin{eqnarray}
&&
\left\langle
\mathrm{P}_{\nu_{e}\to\nu_{\mathrm{S}}}
\right\rangle_{\mathrm{NC}}
\ge
1
-
\left( \mathrm{R}^{\mathrm{ES}}_{\mathrm{NC}} \right)^{-1}
\; ,
\label{E523}
\\
&&
\Phi
\ge
{\displaystyle
N^{\mathrm{NC}}
\over\displaystyle
X_{{\nu}d}
\,
\left( \mathrm{R}^{\mathrm{ES}}_{\mathrm{NC}} \right)^{-1}
}
=
{\displaystyle
\Sigma^{\mathrm{ES}}
\over\displaystyle
X_{\nu_{\mu}e}
}
\; .
\label{E524}
\end{eqnarray}

\begin{table}[t]
\begin{center}
\begin{tabular*}{\textwidth}
{c@{\extracolsep{\fill}}
 c@{\extracolsep{\fill}}
 c@{\extracolsep{\fill}}
 c@{\extracolsep{\fill}}
 c}
\hline
\hline
\\
$ \nu_{e} \to \nu_{\mathrm{S}} $
&
$ \Delta m^2 \, \mathrm{(eV^2)} $
&
$ \sin^2 2\theta $
&
$
\left[
\mathrm{P}_{\nu_{e}\to\nu_{e}}
\right]_{\mathrm{max}}
$
&
$ \mathrm{R}^{\mathrm{NC}}_{\mathrm{CC}} $
\\
\\
\hline
\\
MSW
&
$ 4.5 \times 10^{-6} $
&
$ 7.0 \times 10^{-3} $
&
0.57
&
0.71
\\
\\
VACUUM OSC.
&
$ 6.3 \times 10^{-11} $
&
$ 0.85 $
&
0.56
&
0.50
\\
\\
\hline
\hline
\end{tabular*}
\end{center}
\protect\caption{\small
Results of the calculation of
$
\left[
\mathrm{P}_{\nu_{e}\to\nu_{e}}
\right]_{\mathrm{max}}
$
and
$ \mathrm{R}^{\mathrm{NC}}_{\mathrm{CC}} $
in the model with
$\nu_e$--$\nu_{\mathrm{S}}$
mixing.
The values of
$\Delta m^2$ and $\sin^2 2\theta$
used are also given.
These values were obtained
from the analysis of the existing experimental data
(Ref.\protect\cite{B:BHKL}
for the MSW transitions
and Ref.\protect\cite{B:KP}
for the vacuum oscillations).}
\label{T:STE}
\end{table}

\begin{figure}[t]
\begin{center}
\mbox{\epsfig{file=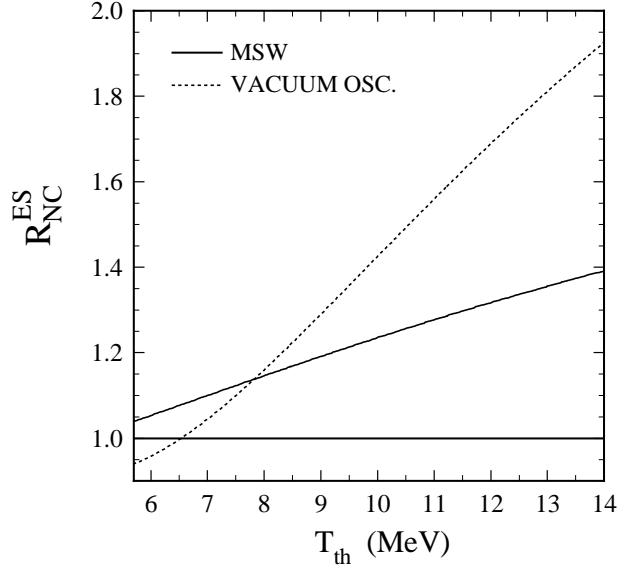,width=0.5\linewidth}}
\end{center}
\protect\caption{\small
Results of the calculation of the ratio
$ \mathrm{R}^{\mathrm{ES}}_{\mathrm{NC}} $
(see Eq.(\protect\ref{E513}))
in the model with
$\nu_{e}$--$\nu_{\mathrm{S}}$
mixing
($\mathrm{T}_{\mathrm{th}}$
is the kinetic energy threshold
in the ES process).
The curves correspond to
MSW transitions
and
vacuum oscillations.
The values of the parameters
$\Delta m^2$ and $\sin^2 2\theta$
used in the calculation
are given in Table~\ref{T:STE}.}
\label{F03}
\end{figure}

It is instructive to calculate
the measurable ratio
$ \mathrm{R}^{\mathrm{ES}}_{\mathrm{NC}} $
(and other ratios that we will consider later)
in some model.
We considered the simplest model
with
$\nu_{e}$--$\nu_{\mathrm{S}}$
mixing.
The parameters of the model
$ \Delta m^2 $ and $ \sin^2 2 \theta $,
which are given in Table~\ref{T:STE},
were obtained from the fit of the existing solar neutrino data
under the assumption of
MSW transitions or vacuum oscillations
(assuming the SSM neutrino fluxes).
The ratio
$ \mathrm{R}^{\mathrm{ES}}_{\mathrm{NC}} $
as a function of the recoil electron threshold energy
is depicted in Fig.\ref{F03}.
{}From Fig.\ref{F03}
it can be seen that
in the case of MSW transitions
the ratio
$ \mathrm{R}^{\mathrm{ES}}_{\mathrm{NC}} $
is bigger than one
for all the considered threshold energies.
In both the MSW and vacuum oscillation cases
the ratio
$ \mathrm{R}^{\mathrm{ES}}_{\mathrm{NC}} $
increases with
$ \mathrm{T}_{\mathrm{th}} $.
{}From Fig.\ref{F03}
it follows that
a high threshold energy
is preferable for revealing the existence
of sterile neutrinos.

\begin{figure}[t]
\begin{minipage}[t]{0.49\linewidth}
\begin{center}
\mbox{\epsfig{file=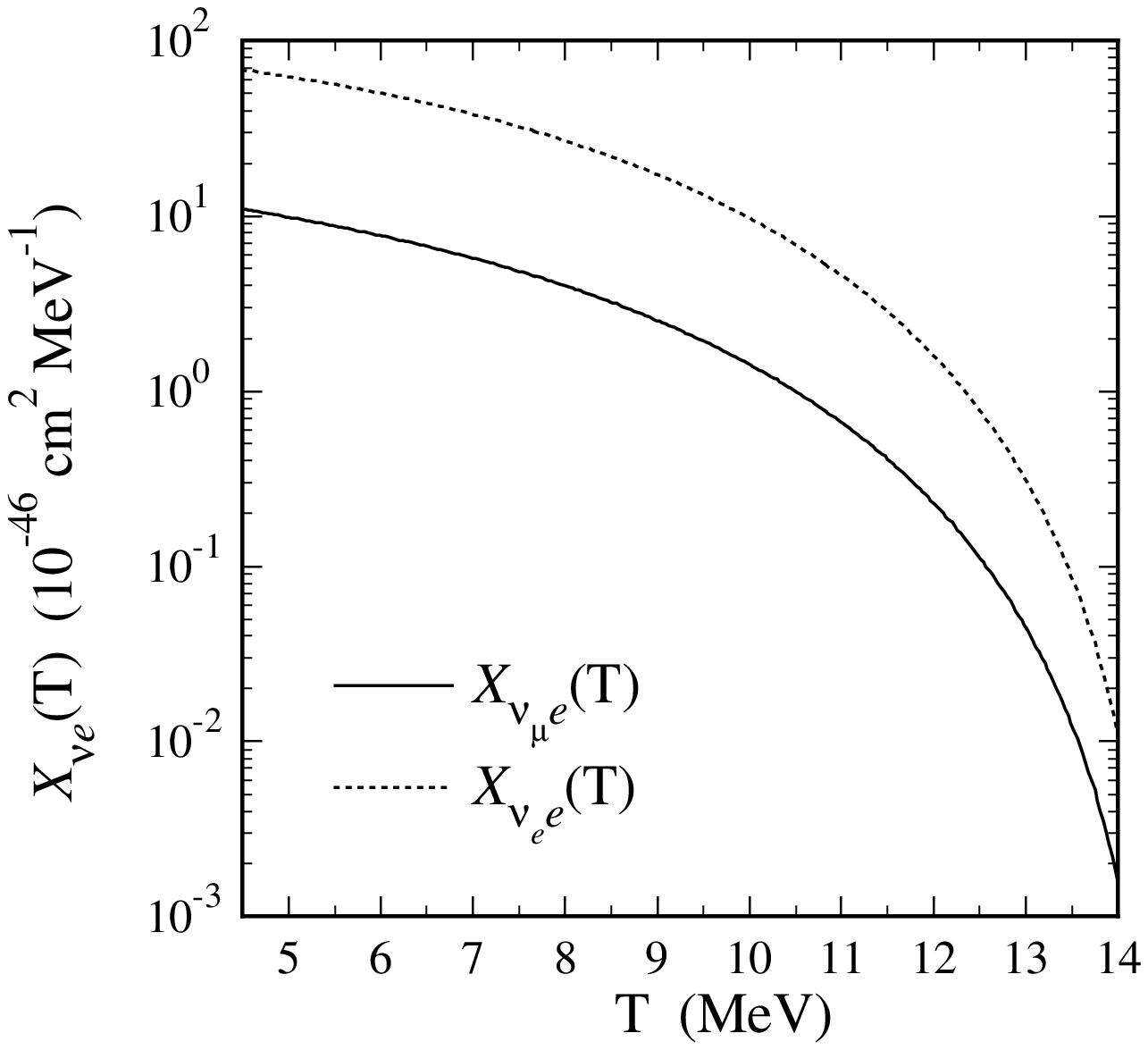,width=\linewidth}}
\end{center}
\protect\caption{\small
Plot of the functions
$ X_{\nu_{\mu}e}(\mathrm{T}) $
and
$ X_{\nu_{e}e}(\mathrm{T}) $
defined in
Eqs.(\protect\ref{E533}) and (\protect\ref{E575}),
respectively.
The depicted range for
the kinetic energy
$\mathrm{T}$
of the recoil electrons in the ES process
will be explored by SNO
with $\mathrm{T}_{\mathrm{th}}=4.5\,\mathrm{MeV}$.}
\label{F04}
\end{minipage}
\hfill
\begin{minipage}[t]{0.49\linewidth}
\begin{center}
\mbox{\epsfig{file=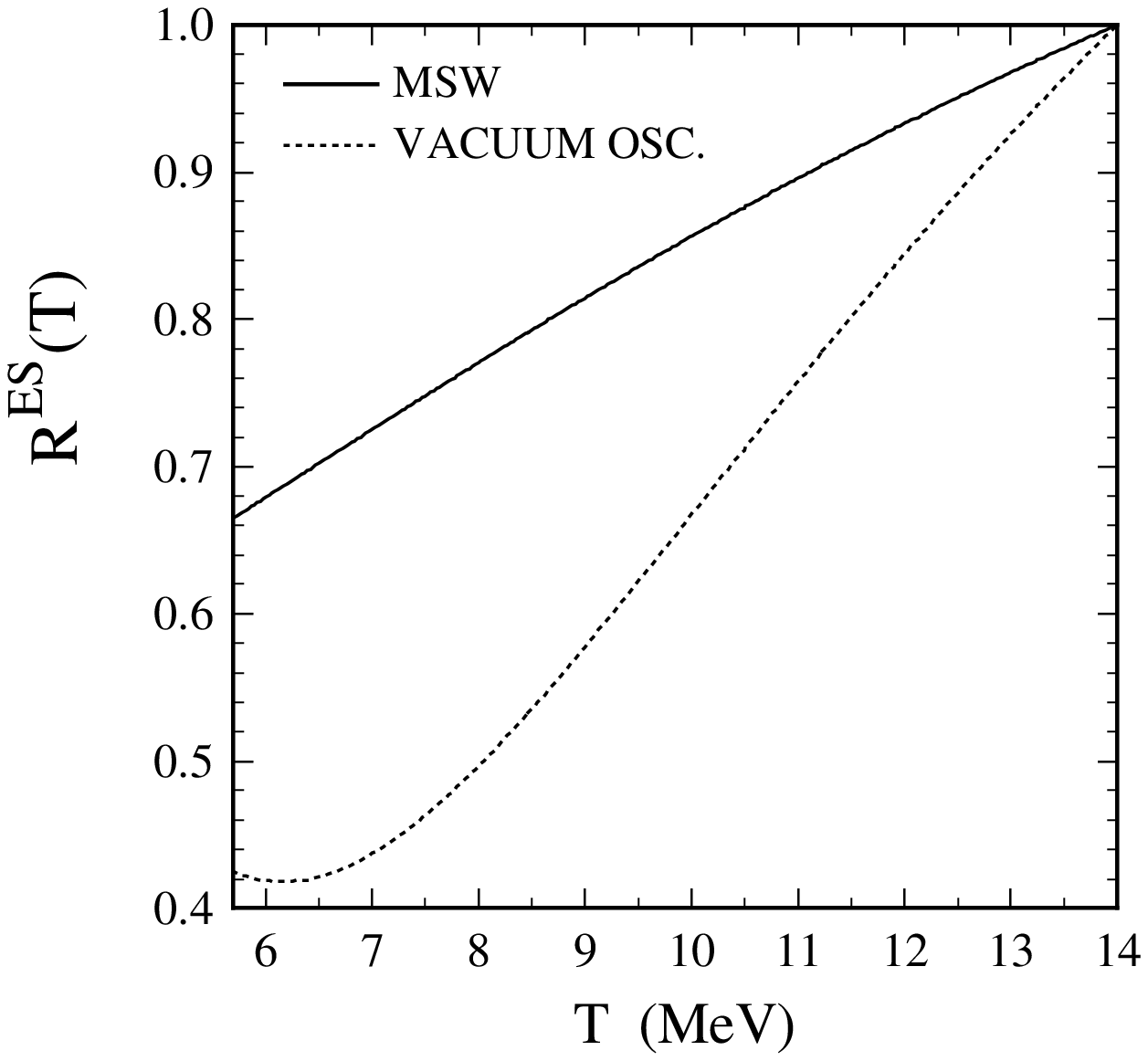,width=\linewidth}}
\end{center}
\protect\caption{\small
Results of the calculation of the ratio
$ \mathrm{R}^{\mathrm{ES}}(\mathrm{T}) $
(see Eq.(\protect\ref{E537}))
in the model with
$\nu_{e}$--$\nu_{\mathrm{S}}$
mixing
($\mathrm{T}$
is the kinetic energy of the recoil electrons in the ES process).
The curves correspond to
MSW transitions
and
vacuum oscillations.
The values of the parameters
$\Delta m^2$ and $\sin^2 2\theta$
used in the calculation
are given in Table~\ref{T:STE}.}
\label{F05}
\end{minipage}
\end{figure}

A detailed investigation
of the spectrum of the recoil electrons
in the process $ \nu e \to \nu e $
will be carried out
in the S-K and SNO experiments.
We will discuss now
what additional model independent tests
of the existence of sterile neutrinos
can be performed
when the recoil electron spectrum will be available.
The spectrum of recoil electrons is given by
\begin{eqnarray}
n^{\mathrm{ES}}(\mathrm{T})
& = &
\int_{E_{\mathrm{m}}(\mathrm{T})}
{\displaystyle
\mathrm{d} \sigma_{\nu_{e}e}
\over\displaystyle
\mathrm{d} \mathrm{T}
}
(E,\mathrm{T})
\,
\mathrm{P}_{\nu_{e}\to\nu_{e}}(E)
\,
X(E)
\,
\mathrm{d} E
\,
\Phi
\nonumber
\\
&&
\mbox{}
+
\int_{E_{\mathrm{m}}(\mathrm{T})}
{\displaystyle
\mathrm{d} \sigma_{\nu_{\mu}e}
\over\displaystyle
\mathrm{d} \mathrm{T}
}
(E,\mathrm{T})
\,
\sum_{\ell=\mu,\tau} \mathrm{P}_{\nu_{e}\to\nu_{\ell}}(E)
\,
X(E)
\,
\mathrm{d} E
\,
\Phi
\; .
\label{E531}
\end{eqnarray}
Here $\mathrm{T}$
is the kinetic energy of the recoil electrons,
$ \displaystyle
E_{\mathrm{m}}(\mathrm{T})
=
\mathrm{T}
\left(
1
+
\sqrt{ 1 + 2 \, m_{e} / \mathrm{T} }
\right)
\big/ 2
$
and
$ \displaystyle
{\displaystyle
\mathrm{d} \sigma_{\nu_{\ell}e}
\over\displaystyle
\mathrm{d} \mathrm{T}
}
(E,\mathrm{T})
$
is the differential cross section
of the process
$ \nu_{\ell} e \to \nu_{\ell} e $
($\ell=e,\mu$).
With the help of Eq.(\ref{E531})
we get the following relation:
\begin{equation}
\left\langle
\sum_{\ell=e,\mu,\tau} \mathrm{P}_{\nu_{e}\to\nu_{\ell}}
\right\rangle_{\mathrm{ES};\mathrm{T}}
=
{\displaystyle
\Sigma^{\mathrm{ES}}(\mathrm{T})
\over\displaystyle
\Phi
\,
X_{\nu_{\mu}e}(\mathrm{T})
}
\; .
\label{E532}
\end{equation}
Here
\begin{equation}
X_{\nu_{\mu}e}(\mathrm{T})
\equiv
\int_{E_{\mathrm{m}}(\mathrm{T})}
{\displaystyle
\mathrm{d} \sigma_{\nu_{\mu}e}
\over\displaystyle
\mathrm{d} \mathrm{T}
}
(E,\mathrm{T})
\,
X(E)
\,
\mathrm{d} E
\label{E533}
\end{equation}
is a known function,
which is plotted in Fig.\ref{F04}.
Other quantities in Eq.(\ref{E532})
are determined as follows:
\begin{equation}
\Sigma^{\mathrm{ES}}(\mathrm{T})
\equiv
n^{\mathrm{ES}}(\mathrm{T})
-
\int_{E_{\mathrm{m}}(\mathrm{T})}
\left[
{\displaystyle
\mathrm{d} \sigma_{\nu_{e}e}
\over\displaystyle
\mathrm{d} \mathrm{T}
}
(E,\mathrm{T})
-
{\displaystyle
\mathrm{d} \sigma_{\nu_{\mu}e}
\over\displaystyle
\mathrm{d} \mathrm{T}
}
(E,\mathrm{T})
\right]
\phi_{\nu_{e}}(E)
\,
\mathrm{d} E
\label{E534}
\end{equation}
and
\begin{equation}
\left\langle
\sum_{\ell=e,\mu,\tau} \mathrm{P}_{\nu_{e}\to\nu_{\ell}}
\right\rangle_{\mathrm{ES};\mathrm{T}}
\equiv
{\displaystyle
1
\over\displaystyle
X_{\nu_{\mu}e}(\mathrm{T})
}
\int_{E_{\mathrm{m}}(\mathrm{T})}
{\displaystyle
\mathrm{d} \sigma_{\nu_{\mu}e}
\over\displaystyle
\mathrm{d} \mathrm{T}
}
(E,\mathrm{T})
\,
X(E)
\sum_{\ell=e,\mu,\tau} \mathrm{P}_{\nu_{e}\to\nu_{\ell}}(E) \,
\mathrm{d} E
\; .
\label{E535}
\end{equation}
Let us stress that to determine the quantity
$ \Sigma^{\mathrm{ES}}(\mathrm{T}) $
it is necessary to know
the recoil electron energy spectrum
$ n^{\mathrm{ES}}(\mathrm{T}) $
as well as the spectrum of $\nu_e$
on the earth $ \phi_{\nu_{e}}(E) $.

Let us consider now the relation (\ref{E532}).
If the quantity
$ \displaystyle
\Sigma^{\mathrm{ES}}(\mathrm{T})
/
X_{\nu_{\mu}e}(\mathrm{T})
$
depends on the energy $\mathrm{T}$,
it would mean that
transitions
of solar $\nu_e$'s
into sterile neutrinos take place
(if there are no such transitions,
then
$ \displaystyle
\left\langle
\sum_{\ell=e,\mu,\tau} \mathrm{P}_{\nu_{e}\to\nu_{\ell}}
\right\rangle_{\mathrm{ES};\mathrm{T}}
=
1
$
and
$ \displaystyle
\Sigma^{\mathrm{ES}}(\mathrm{T})
/
X_{\nu_{\mu}e}(\mathrm{T})
=
\mbox{const}
$).
In order to derive
a lower bound for the average value
of the probability of transition
of $\nu_e$'s into sterile states
we will use the following inequality:
\begin{equation}
\left\langle
\sum_{\ell=e,\mu,\tau} \mathrm{P}_{\nu_{e}\to\nu_{\ell}}
\right\rangle_{\mathrm{ES};\mathrm{T}}
\le
{\displaystyle
\left\langle
\sum_{\ell=e,\mu,\tau} \mathrm{P}_{\nu_{e}\to\nu_{\ell}}
\right\rangle_{\mathrm{ES};\mathrm{T}}
\over\displaystyle
\left[
\left\langle
\sum_{\ell=e,\mu,\tau} \mathrm{P}_{\nu_{e}\to\nu_{\ell}}
\right\rangle_{\mathrm{ES};\mathrm{T}}
\right]_{\mathrm{max}}
}
=
\mathrm{R}^{\mathrm{ES}}(\mathrm{T})
\;.
\label{E536}
\end{equation}
Here
\begin{equation}
\mathrm{R}^{\mathrm{ES}}(\mathrm{T})
\equiv
{\displaystyle
\Sigma^{\mathrm{ES}}(\mathrm{T})
/
X_{\nu_{\mu}e}(\mathrm{T})
\over\displaystyle
\left[
\Sigma^{\mathrm{ES}}(\mathrm{T})
/
X_{\nu_{\mu}e}(\mathrm{T})
\right]_{\mathrm{max}}
}
\label{E537}
\end{equation}
is a measurable quantity
and the subscript max
indicates the maximum value
in the explored energy range.
{}From Eq.(\ref{E536})
it follows that
\begin{equation}
\left\langle
\mathrm{P}_{\nu_{e}\to\nu_{\mathrm{S}}}
\right\rangle_{\mathrm{ES};\mathrm{T}}
\ge
1
-
\mathrm{R}^{\mathrm{ES}}(\mathrm{T})
\; .
\label{E538}
\end{equation}

For the total initial flux
of $^8\mathrm{B}$ neutrinos,
from Eqs.(\ref{E532}) and (\ref{E536})
we obtain the following inequality:
\begin{equation}
\Phi
\ge
\left[
{\displaystyle
\Sigma^{\mathrm{ES}}(\mathrm{T})
\over\displaystyle
X_{\nu_{\mu}e}(\mathrm{T})
}
\right]_{\mathrm{max}}
\; .
\label{E539}
\end{equation}

The results of the calculation of the ratio
$ \mathrm{R}^{\mathrm{ES}}(\mathrm{T}) $
in the model with
$\nu_{e}$--$\nu_{\mathrm{S}}$
mixing
is presented in Fig.\ref{F05}.
{}From this figure it is seen
that a detailed investigation
of the ratio
$ \mathrm{R}^{\mathrm{ES}}(\mathrm{T}) $
near the threshold
could be a promising way
to search for
$\nu_{e}\to\nu_{\mathrm{S}}$
transitions.

We will obtain now other inequalities
the test of which could allow
to obtain model independent informations
about the existence of sterile neutrinos.
We have
\begin{equation}
\left\langle
\sum_{\ell=e,\mu,\tau} \mathrm{P}_{\nu_{e}\to\nu_{\ell}}
\right\rangle_{\mathrm{ES};\mathrm{T}}
\le
{\displaystyle
\left\langle
\sum_{\ell=e,\mu,\tau} \mathrm{P}_{\nu_{e}\to\nu_{\ell}}
\right\rangle_{\mathrm{ES};\mathrm{T}}
\over\displaystyle
\left\langle
\sum_{\ell=e,\mu,\tau} \mathrm{P}_{\nu_{e}\to\nu_{\ell}}
\right\rangle_{a}
}
=
\mathrm{R}^{\mathrm{ES}}_{a}(\mathrm{T})
\; .
\label{E540}
\end{equation}
Here
$ a = \mathrm{NC} , \, \mathrm{ES} $
and the ratios
\begin{eqnarray}
&&
\mathrm{R}^{\mathrm{ES}}_{\mathrm{NC}}(\mathrm{T})
\equiv
{\displaystyle
\Sigma^{\mathrm{ES}}(\mathrm{T})
\,
X_{{\nu}d}
\over\displaystyle
X_{\nu_{\mu}e}(\mathrm{T})
\,
N^{\mathrm{NC}}
}
\; ,
\label{E541}
\\
&&
\mathrm{R}^{\mathrm{ES}}_{\mathrm{ES}}(\mathrm{T})
\equiv
{\displaystyle
\Sigma^{\mathrm{ES}}(\mathrm{T})
\,
X_{\nu_{\mu}e}
\over\displaystyle
X_{\nu_{\mu}e}(\mathrm{T})
\,
\Sigma^{\mathrm{ES}}
}
\label{E542}
\end{eqnarray}
are measurable quantities.

If the inequality
\begin{equation}
\mathrm{R}^{\mathrm{ES}}_{\mathrm{NC}}(\mathrm{T})
<
1
\label{E543}
\end{equation}
takes place in some region of $\mathrm{T}$,
it would mean that
there are transitions of solar $\nu_e$'s
into sterile states.
{}From Eqs.(\ref{E532}) and (\ref{E540})
we find the following lower bounds
for
$ \displaystyle
\left\langle
\mathrm{P}_{\nu_{e}\to\nu_{\mathrm{S}}}
\right\rangle_{\mathrm{ES};\mathrm{T}}
$
and for the initial total $\nu_{e}$ flux:
\begin{eqnarray}
&&
\left\langle
\mathrm{P}_{\nu_{e}\to\nu_{\mathrm{S}}}
\right\rangle_{\mathrm{ES};\mathrm{T}}
\ge
1
-
\mathrm{R}^{\mathrm{ES}}_{\mathrm{NC}}(\mathrm{T})
\; ,
\label{E545}
\\
&&
\Phi
\ge
{\displaystyle
N^{\mathrm{NC}}
\over\displaystyle
X_{{\nu}d}
}
\; .
\label{E546}
\end{eqnarray}
Analogously,
if in some region of $ (\mathrm{T}) $
the inequality
\begin{equation}
\mathrm{R}^{\mathrm{ES}}_{\mathrm{ES}}(\mathrm{T})
<
1
\label{E544}
\end{equation}
is satisfied,
we have the following lower bounds
for
$ \displaystyle
\left\langle
\mathrm{P}_{\nu_{e}\to\nu_{\mathrm{S}}}
\right\rangle_{\mathrm{ES};\mathrm{T}}
$
and
$ \Phi $:
\begin{eqnarray}
&&
\left\langle
\mathrm{P}_{\nu_{e}\to\nu_{\mathrm{S}}}
\right\rangle_{\mathrm{ES};\mathrm{T}}
\ge
1
-
\mathrm{R}^{\mathrm{ES}}_{\mathrm{ES}}(\mathrm{T})
\; ,
\label{E547}
\\
&&
\Phi
\ge
{\displaystyle
\Sigma^{\mathrm{ES}}
\over\displaystyle
X_{\nu_{\mu}e}
}
\; .
\label{E548}
\end{eqnarray}

Further,
we have
\begin{equation}
\left\langle
\sum_{\ell=e,\mu,\tau} \mathrm{P}_{\nu_{e}\to\nu_{\ell}}
\right\rangle_{a}
\le
{\displaystyle
\left\langle
\sum_{\ell=e,\mu,\tau} \mathrm{P}_{\nu_{e}\to\nu_{\ell}}
\right\rangle_{a}
\over\displaystyle
\left\langle
\sum_{\ell=e,\mu,\tau} \mathrm{P}_{\nu_{e}\to\nu_{\ell}}
\right\rangle_{\mathrm{ES};\mathrm{T}}
}
=
{\displaystyle
1
\over\displaystyle
\mathrm{R}^{\mathrm{ES}}_{a}(\mathrm{T})
}
\; ,
\label{E551}
\end{equation}
where
$ a = \mathrm{NC} , \, \mathrm{ES} $.
Thus,
sterile neutrinos exist if
in some region of the variable
$ \mathrm{T} $
the inequality
\begin{equation}
\mathrm{R}^{\mathrm{ES}}_{a}(\mathrm{T})
>
1
\label{E552}
\end{equation}
is satisfied.
For the averaged probability of the transition
of $\nu_e$ into $\nu_{\mathrm{S}}$
and for the total flux we have
\begin{eqnarray}
&&
\left\langle
\mathrm{P}_{\nu_{e}\to\nu_{\mathrm{S}}}
\right\rangle_{a}
\ge
1
-
\left ( \mathrm{R}^{\mathrm{ES}}_{a}(\mathrm{T}) \right)_{\mathrm{max}}^{-1}
\; ,
\label{E553}
\\
&&
\Phi
\ge
\left(
{\displaystyle
\Sigma^{\mathrm{ES}}(\mathrm{T})
\over\displaystyle
X_{\nu_{\mu}e}(\mathrm{T})
}
\right)_{\mathrm{max}}
\; .
\label{E554}
\end{eqnarray}

\begin{figure}[t]
\begin{minipage}[t]{0.49\linewidth}
\begin{center}
\mbox{\epsfig{file=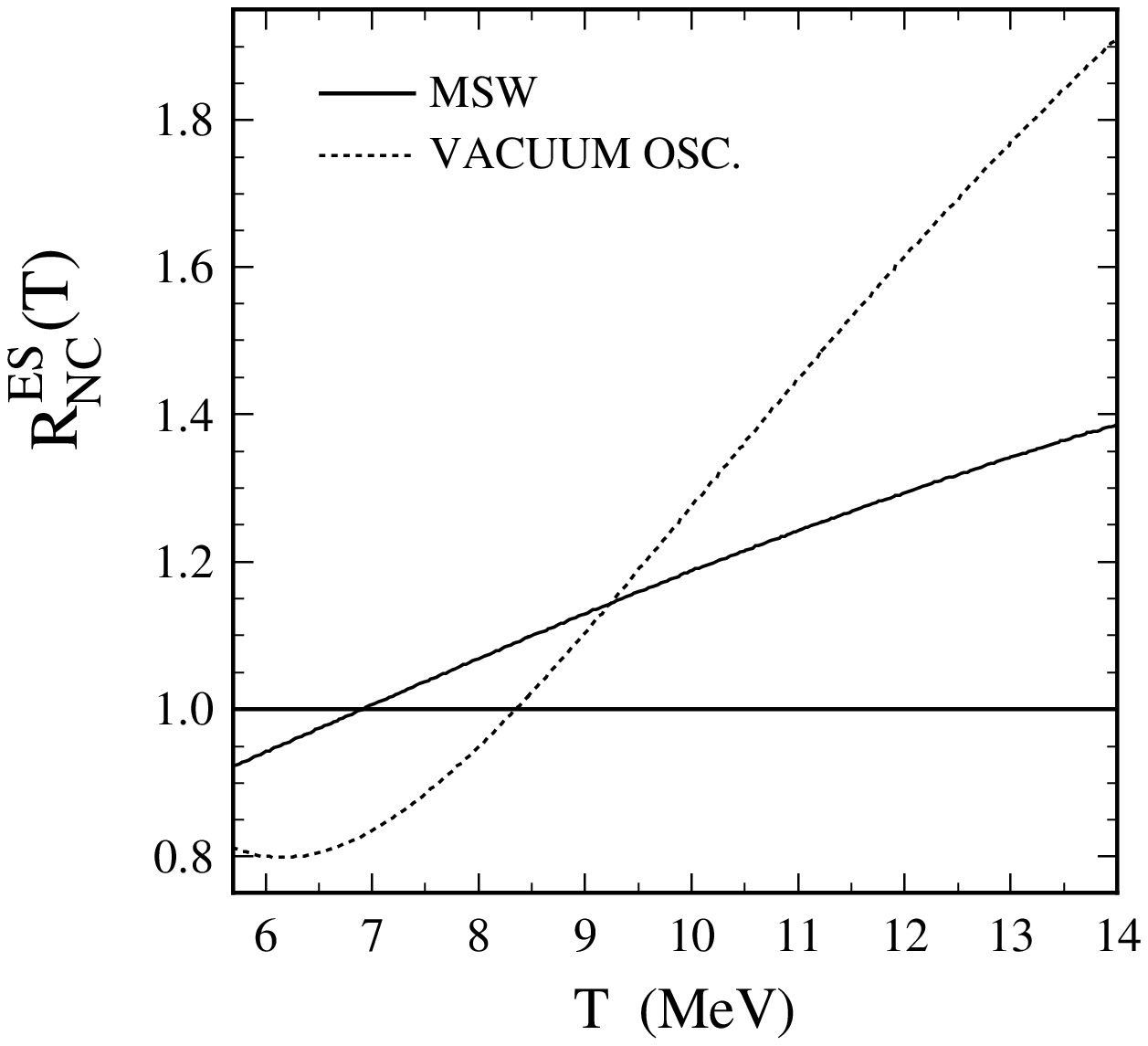,width=\linewidth}}
\end{center}
\protect\caption{\small
Results of the calculation of the ratio
$ \mathrm{R}^{\mathrm{ES}}_{\mathrm{NC}}(\mathrm{T}) $
(see Eq.(\protect\ref{E541}))
in the model with
$\nu_{e}$--$\nu_{\mathrm{S}}$
mixing.}
\label{F06}
\end{minipage}
\hfill
\begin{minipage}[t]{0.49\linewidth}
\begin{center}
\mbox{\epsfig{file=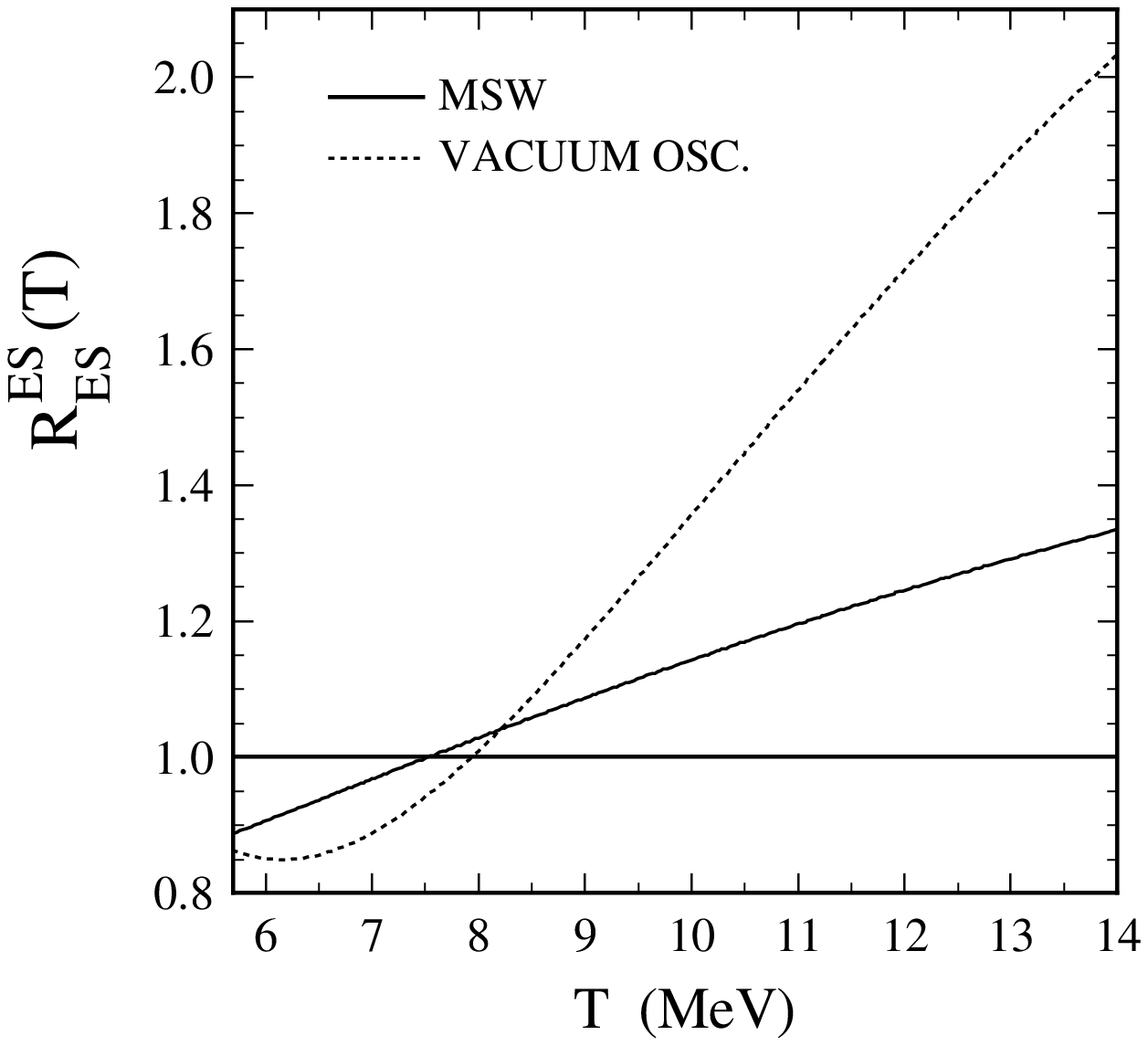,width=\linewidth}}
\end{center}
\protect\caption{\small
Results of the calculation of the ratio
$ \mathrm{R}^{\mathrm{ES}}_{\mathrm{ES}}(\mathrm{T}) $
(see Eq.(\protect\ref{E542}))
in the model with
$\nu_{e}$--$\nu_{\mathrm{S}}$
mixing.}
\label{F07}
\end{minipage}
\end{figure}

In Figs.\ref{F06} and \ref{F07}
we plotted the ratios
$ \mathrm{R}^{\mathrm{ES}}_{NC}(\mathrm{T}) $
and
$ \mathrm{R}^{\mathrm{ES}}_{ES}(\mathrm{T}) $
calculated in the model with
$\nu_{e}$--$\nu_{\mathrm{S}}$
mixing
(the parameters of the model are given in Table~\ref{T:STE}).
It can be seen from these figures
that in the model under consideration
the ratios
$ \mathrm{R}^{\mathrm{ES}}_{NC}(\mathrm{T}) $
and
$ \mathrm{R}^{\mathrm{ES}}_{ES}(\mathrm{T}) $
are larger than one
in a large part of the
kinematical region.
The deviation of
$ \mathrm{R}^{\mathrm{ES}}_{NC}(\mathrm{T}) $
and
$ \mathrm{R}^{\mathrm{ES}}_{ES}(\mathrm{T}) $
from one
is larger
in the case of vacuum oscillations
than in the MSW case.

We have derived several different inequalities
whose test could allow
to reveal the presence of sterile neutrinos
in the solar neutrino flux on the earth
in a model independent way.
Additional inequalities of this type
can be obtained
from the knowledge of
the spectrum
of solar neutrinos on the earth
$ \phi_{\nu_{e}}(E) $.
The relation
\begin{equation}
\mathrm{P}_{\nu_{e}\to\nu_{e}}(E)
=
{\displaystyle
\phi_{\nu_{e}}(E)
\over\displaystyle
\Phi
\,
X(E)
}
\label{E555}
\end{equation}
and the relations
(\ref{E503}), (\ref{E509}) and (\ref{E532})
have a similar structure.
We have
\begin{equation}
\left\langle
\sum_{\ell=e,\mu,\tau} \mathrm{P}_{\nu_{e}\to\nu_{\ell}}
\right\rangle_{a}
\le
{\displaystyle
\left\langle
\sum_{\ell=e,\mu,\tau} \mathrm{P}_{\nu_{e}\to\nu_{\ell}}
\right\rangle_{a}
\over\displaystyle
\left[
\mathrm{P}_{\nu_{e}\to\nu_{e}}
\right]_{\mathrm{max}}
}
=
\mathrm{R}^{a}_{\mathrm{CC}}
\; ,
\label{E556}
\end{equation}
where
$ a = \mathrm{NC} , \, \mathrm{ES} $
and the ratios
\begin{eqnarray}
&&
\mathrm{R}^{\mathrm{NC}}_{\mathrm{CC}}
\equiv
{\displaystyle
N^{\mathrm{NC}}
/
X_{{\nu}d}
\over\displaystyle
\left[
\phi_{\nu_e}
/
X
\right]_{\mathrm{max}}
}
\; ,
\label{E557}
\\
&&
\mathrm{R}^{\mathrm{ES}}_{\mathrm{CC}}
\equiv
{\displaystyle
\Sigma^{\mathrm{ES}}
/
X_{\nu_{\mu}e}
\over\displaystyle
\left[
\phi_{\nu_e}
/
X
\right]_{\mathrm{max}}
}
\label{E558}
\end{eqnarray}
are measurable quantities.
The quantity
$ \displaystyle
\left[
\phi_{\nu_e}
/
X
\right]_{\mathrm{max}}
$
in Eq.(\ref{E557}) and (\ref{E558})
is the maximal value
of the function
$ \displaystyle
\phi_{\nu_e}(E)
/
X(E)
$
in the explored energy range.
{}From Eq.(\ref{E556}),
for the averaged values of the probability
$ \mathrm{P}_{\nu_{e}\to\nu_{\mathrm{S}}} $
and for the total flux
$ \Phi $
we have
\begin{eqnarray}
&&
\left\langle
\mathrm{P}_{\nu_{e}\to\nu_{\mathrm{S}}}
\right\rangle_{a}
\ge
1
-
\mathrm{R}^{a}_{\mathrm{CC}}
\quad \quad
( a = \mathrm{NC} , \, \mathrm{ES} )
\; ,
\label{E559}
\\
&&
\Phi
\ge
\left[
{\displaystyle
\phi_{\nu_e}
\over\displaystyle
X
}
\right]_{\mathrm{max}}
\; .
\label{E560}
\end{eqnarray}

\begin{figure}[t]
\begin{minipage}[t]{0.49\linewidth}
\begin{center}
\mbox{\epsfig{file=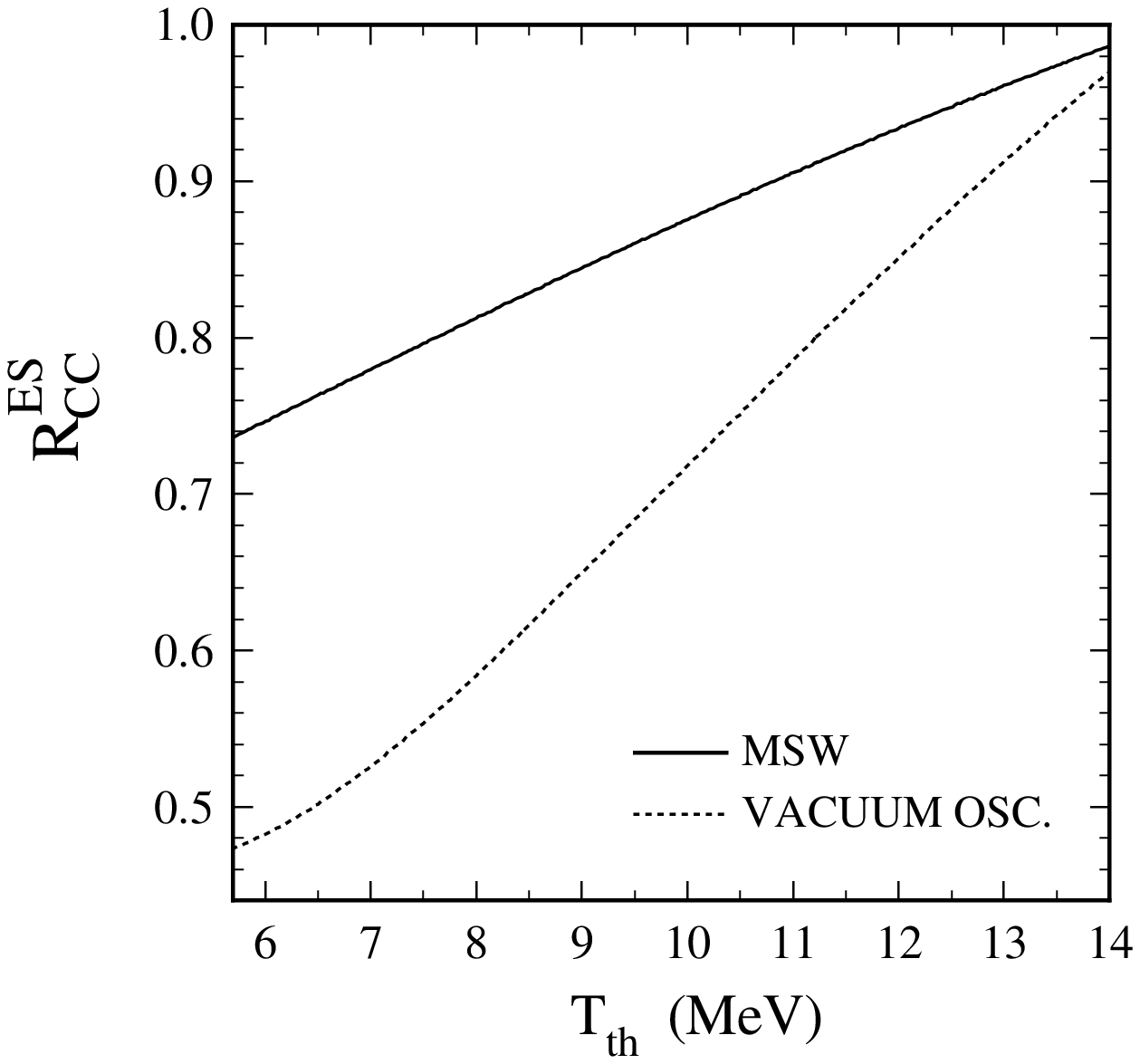,width=\linewidth}}
\end{center}
\protect\caption{\small
Results of the calculation of the ratio
$ \mathrm{R}^{\mathrm{ES}}_{\mathrm{CC}} $
(see Eq.(\protect\ref{E558}))
in the model with
$\nu_{e}$--$\nu_{\mathrm{S}}$
mixing
($\mathrm{T}_{\mathrm{th}}$
is the kinetic energy threshold
of the recoil electrons in the ES process).
The curves correspond to
MSW transitions
and
vacuum oscillations.}
\label{F08}
\end{minipage}
\hfill
\begin{minipage}[t]{0.49\linewidth}
\begin{center}
\mbox{\epsfig{file=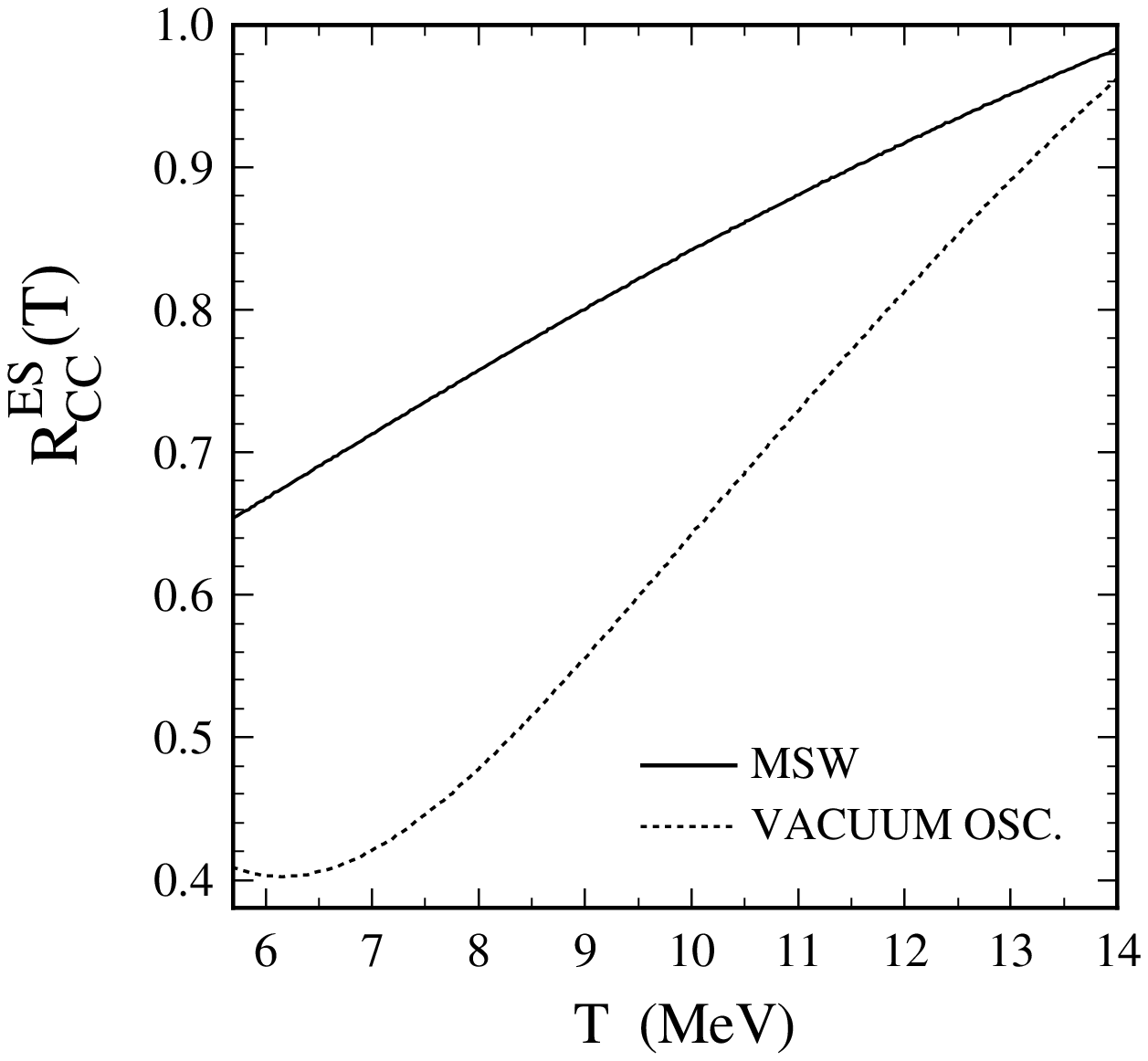,width=\linewidth}}
\end{center}
\protect\caption{\small
Results of the calculation of the ratio
$ \mathrm{R}^{\mathrm{ES}}_{\mathrm{CC}}(\mathrm{T}) $
(see Eq.(\protect\ref{E562}))
in the model with
$\nu_{e}$--$\nu_{\mathrm{S}}$
mixing
($\mathrm{T}$
is the kinetic energy
of the recoil electrons in the ES process).
The curves correspond to
MSW transitions
and
vacuum oscillations.}
\label{F09}
\end{minipage}
\end{figure}

The results of our calculations of the ratio
$ \mathrm{R}^{\mathrm{NC}}_{\mathrm{CC}} $
in the model with
$\nu_{e}$--$\nu_{\mathrm{S}}$
mixing
are given in Table~\ref{T:STE},
whereas the results of our calculations of the ratio
$ \mathrm{R}^{\mathrm{ES}}_{\mathrm{CC}} $
as a function
of the kinetic energy threshold
of the recoil electrons in the ES process,
$\mathrm{T}_{\mathrm{th}}$,
are depicted in Fig.~\ref{F08}
(the kinetic energy threshold
of the electrons in the CC process
is assumed to be 4.5 MeV).
It can be seen from Fig.~\ref{F08}
that in the model under consideration
the ratio
$ \mathrm{R}^{\mathrm{ES}}_{\mathrm{CC}} $
depends strongly on
$\mathrm{T}_{\mathrm{th}}$
(especially in the case of vacuum oscillations).
It is preferable to search for
$ \nu_e \to \nu_{\mathrm{S}} $
transitions
at relatively small thresholds.

Finally,
from Eqs.(\ref{E532}) and  (\ref{E555})
it follows
\begin{equation}
\left\langle
\sum_{\ell=e,\mu,\tau} \mathrm{P}_{\nu_{e}\to\nu_{\ell}}
\right\rangle_{\mathrm{ES};\mathrm{T}}
\le
\mathrm{R}^{\mathrm{ES}}_{\mathrm{CC}}(\mathrm{T})
\; ,
\label{E561}
\end{equation}
where the ratio
\begin{equation}
\mathrm{R}^{\mathrm{ES}}_{\mathrm{CC}}(\mathrm{T})
\equiv
{\displaystyle
\Sigma^{\mathrm{ES}}(\mathrm{T})
\over\displaystyle
X_{\nu_{\mu}e}(\mathrm{T})
\left[
\phi_{\nu_e}
/
X
\right]_{\mathrm{max}}
}
\label{E562}
\end{equation}
is a measurable quantity.
{}From
Eqs.(\ref{E532}) and (\ref{E561})
we have
\begin{eqnarray}
&&
\left\langle
\mathrm{P}_{\nu_{e}\to\nu_{\mathrm{S}}}
\right\rangle_{\mathrm{ES};\mathrm{T}}
\ge
1
-
\mathrm{R}^{\mathrm{ES}}_{\mathrm{CC}}(\mathrm{T})
\; ,
\label{E563}
\\
&&
\Phi
\ge
\left[
{\displaystyle
\phi_{\nu_e}
\over\displaystyle
X
}
\right]_{\mathrm{max}}
\; .
\label{E564}
\end{eqnarray}
The results of the calculations
of the ratio
$ \mathrm{R}^{\mathrm{ES}}_{\mathrm{CC}}(\mathrm{T}) $
in the model with
$\nu_{e}$--$\nu_{\mathrm{S}}$
mixing
are presented in Fig.\ref{F09}.
In this model there are large deviations
from one
of the ratio
$ \mathrm{R}^{\mathrm{ES}}_{\mathrm{CC}}(\mathrm{T}) $
for values of
$ \mathrm{T} $
close to the threshold
(especially in the case of vacuum oscillations).

In conclusion,
we would like to stress
that all the inequalities
discussed here
are based on the conservation law (\ref{E504}).
The concrete mechanism
for the transition of $\nu_e$'s
into sterile states is not important
for the tests proposed here.
Thus,
there could be
not only
$ \nu_e \to \nu_{\mathrm{S}} $
transitions due to
Dirac-Majorana neutrino mixing,
but also
$ \nu_e \to \nu_{\mathrm{S}} $
transitions due to
neutrino Dirac magnetic moments
\cite{B:SFP}.
In the latter case
the measurable ratios
$ \mathrm{R}^{\mathrm{ES}}_{\mathrm{NC}} $,
$ \mathrm{R}^{\mathrm{ES}}(\mathrm{T}) $,
\ldots
will depend periodically on time
\cite{B:BG94A,B:BG94V}.

\section{The case of absence of
transitions of solar
$ \boldsymbol{\nu_{e}} $'s
into sterile states}

If the test of all the inequalities
obtained in the previous section
will not reveal the presence of
sterile neutrinos in the solar neutrino flux
on the earth,
it will be natural
to consider the possibility
of a model independent treatment of solar neutrino data
under the assumption
that there are no transitions
of solar $\nu_e$'s into sterile states\footnote{
Let us stress that all the inequalities
considered in the previous section
will not reveal the presence of sterile neutrinos
if
$ \mathrm{P}_{\nu_{e}\to\nu_{\mathrm{S}}}(E) = \mbox{const} $.
}.
In this case
\begin{eqnarray}
&&
\left\langle
\sum_{\ell=e,\mu,\tau} \mathrm{P}_{\nu_{e}\to\nu_{\ell}}
\right\rangle_{a}
=
1
\quad \quad
( a = \mathrm{NC} , \, \mathrm{ES} )
\; ,
\label{E565}
\\
&&
\left\langle
\sum_{\ell=e,\mu,\tau} \mathrm{P}_{\nu_{e}\to\nu_{\ell}}
\right\rangle_{\mathrm{ES};\mathrm{T}}
=
1
\; ,
\label{E566}
\end{eqnarray}
and
the SNO and S-K data will allow to
determine the initial flux
of $^8\mathrm{B}$ $\nu_e$'s
with {\em three} methods:

\begin{enumerate}

\item
By means of the measurement
of the NC event rate
$N^{\mathrm{NC}}$.
In fact,
from Eqs.(\ref{E503}) and (\ref{E565})
we have
\begin{equation}
\Phi
=
{\displaystyle
N^{\mathrm{NC}}
\over\displaystyle
X_{{\nu}d}
}
\; ,
\label{E567}
\end{equation}
where
$ X_{{\nu}d} $
is given by Eq.(\ref{E599}).

\item
By means of the measurement
of the total number of ES events
$ N^{\mathrm{ES}} $
and the electron spectrum
in the CC process
(from which the spectrum
of $\nu_e$ on the earth,
$ \phi_{\nu_{e}}(E) $,
will be determined).
{}From Eqs.(\ref{E509}) and (\ref{E565})
we have
\begin{equation}
\Phi
=
{\displaystyle
\Sigma^{\mathrm{ES}}
\over\displaystyle
X_{\nu_{\mu}e}
}
\; ,
\label{E568}
\end{equation}
where
$ \Sigma^{\mathrm{ES}} $
is given by Eq.(\ref{E508})
and
$ X_{\nu_{\mu}e} $
by Eq.(\ref{E510}).

\item
By means of the measurement
of the recoil electron spectrum
$ n^{\mathrm{ES}}(\mathrm{T}) $
in the ES process
and the electron spectrum
in the CC process.
In fact,
from Eqs.(\ref{E532}) and (\ref{E566})
we have
\begin{equation}
\Phi
=
{\displaystyle
\Sigma^{\mathrm{ES}}(\mathrm{T})
\over\displaystyle
X_{\nu_{\mu}e}(\mathrm{T})
}
\; ,
\label{E569}
\end{equation}
where
$ \Sigma^{\mathrm{ES}}(\mathrm{T}) $
is given by Eq.(\ref{E534})
and
$ X_{\nu_{\mu}e}(\mathrm{T}) $
by Eq.(\ref{E533}).

\end{enumerate}

All these different methods
of determination of the total flux
$ \Phi $
will give results which must be in agreement
with each other
(otherwise the ratios
$ \mathrm{R}^{\mathrm{ES}}_{\mathrm{NC}} $,
$ \mathrm{R}^{\mathrm{ES}}(\mathrm{T}) $,
\ldots
considered in the previous section
would be different from one).
Let us stress
that the proposed methods
for the determination of the flux of
$^8\mathrm{B}$ neutrinos
do not depend on the
mechanism of transition
of solar $\nu_e$'s
into other active states.
It is evident that
a comparison of the flux
$ \Phi $
determined directly from the experimental data
with the $^8\mathrm{B}$ flux
predicted by solar models
will be an important test of these models.

The SNO and S-K experiments
will allow also to determine
in a model independent way
the probability of $\nu_e$'s to survive,
$ \mathrm{P}_{\nu_{e}\to\nu_{e}}(E) $.
In fact,
we have
\begin{equation}
\mathrm{P}_{\nu_{e}\to\nu_{e}}(E)
=
{\displaystyle
\phi_{\nu_{e}}(E)
\over\displaystyle
\Phi
\,
X(E)
}
\; ,
\label{E570}
\end{equation}
where
$ \phi_{\nu_{e}}(E) $
is the flux of solar $\nu_e$'s
on the earth
and
the total flux
$ \Phi $
is given by
Eqs.(\ref{E567}), (\ref{E568}) and (\ref{E569}).
If it will occur
that the probability
$ \mathrm{P}_{\nu_{e}\to\nu_{e}}(E) $
is less than one,
this will be a model independent proof
that $\nu_e$'s transform into other states.
A detailed investigation
of the dependence on $E$
of the probability
$ \mathrm{P}_{\nu_{e}\to\nu_{e}}(E) $
will allow to distinguish different mechanisms
of neutrino transitions
(MSW, just-so vacuum oscillations and others)
and to determine the corresponding parameters.

\begin{table}[t]
\begin{center}
\begin{tabular*}{\textwidth}
{c@{\extracolsep{\fill}}
 c@{\extracolsep{\fill}}
 c@{\extracolsep{\fill}}
 c}
\hline
\hline
\\
$ \nu_{e} \to \nu_{\mu(\tau)} $
&
$ \Delta m^2 \, \mathrm{(eV^2)} $
&
$ \sin^2 2\theta $
&
$ \displaystyle
\left\langle
\mathrm{P}_{\nu_{e}\to\nu_{e}}
\right\rangle_{\mathrm{ES}}
$
\\
\\
\hline
\\
SMALL MIX MSW
&
$ 6.1 \times 10^{-6} $
&
$ 6.5 \times 10^{-3} $
&
0.32
\\
\\
LARGE MIX MSW
&
$ 9.4 \times 10^{-6} $
&
$ 0.62 $
&
0.19
\\
\\
VACUUM OSC.
&
$ 8.0 \times 10^{-11} $
&
$ 0.80 $
&
0.31
\\
\\
\hline
\hline
\end{tabular*}
\end{center}
\protect\caption{\small
Results of the calculations of the quantity
$ \displaystyle
\left\langle
\mathrm{P}_{\nu_{e}\to\nu_{e}}
\right\rangle_{\mathrm{ES}}
$
in the simplest model with
$\nu_{e}$--$\nu_{\mu(\tau)}$
mixing.
The values of
$\Delta m^2$ and $\sin^2 2\theta$
used are also given.
These values were obtained
from the analysis of the existing experimental data
(Ref.\protect\cite{B:BHKL}
for the MSW transitions
and Ref.\protect\cite{B:KP}
for the vacuum oscillations).}
\label{T:MSW}
\end{table}

The detection of solar neutrinos
through the observation of
$ \nu e \to \nu e $
scattering
will allow to obtain
additional informations about the $\nu_e$
survival probability.
Let us define the average value
of the probability of $\nu_e$
to survive
$ \displaystyle
\left\langle
\mathrm{P}_{\nu_{e}\to\nu_{e}}
\right\rangle_{\mathrm{ES}}
$
as
\begin{equation}
\left\langle
\mathrm{P}_{\nu_{e}\to\nu_{e}}
\right\rangle_{\mathrm{ES}}
\equiv
{\displaystyle
1
\over\displaystyle
X_{\nu_{e}e}
-
X_{\nu_{\mu}e}
}
\int_{E_{\mathrm{th}}^{\mathrm{ES}}}
\left[
\sigma_{\nu_{e}e}(E)
-
\sigma_{\nu_{\mu}e}(E)
\right]
\,
X(E)
\mathrm{P}_{\nu_{e}\to\nu_{e}}(E)
\,
\mathrm{d} E
\; ,
\label{E571}
\end{equation}
where
\begin{equation}
X_{\nu_{e}e}
\equiv
\int_{E_{\mathrm{th}}^{\mathrm{ES}}}
\sigma_{\nu_{e}e}(E)
\,
X(E)
\,
{\mathrm{d}} E
\label{E572}
\end{equation}
and
$ X_{\nu_{\mu}e} $
is given by Eq.(\ref{E510})
(for
$ E_{\mathrm{th}}^{\mathrm{ES}} = 4.74 \, \mathrm{MeV} $,
which corresponds to a
kinetic energy threshold
$ \mathrm{T}_{\mathrm{th}} = 4.5 \, \mathrm{MeV} $
for the recoil electrons,
we have
$ \displaystyle
X_{\nu_{\mu}e}
=
2.12 \times 10^{-44} \, \mathrm{cm}^2
$
and
$ \displaystyle
X_{\nu_{\mu}e}
=
3.23 \times 10^{-45} \, \mathrm{cm}^2
$).

The quantity
$ \displaystyle
\left\langle
\mathrm{P}_{\nu_{e}\to\nu_{e}}
\right\rangle_{\mathrm{ES}}
$
is connected with the total number of ES events
$ N^{\mathrm{ES}} $
and the total flux
$ \Phi $
by the following relation:
\begin{equation}
\left\langle
\mathrm{P}_{\nu_{e}\to\nu_{e}}
\right\rangle_{\mathrm{ES}}
=
{\displaystyle
1
\over\displaystyle
X_{\nu_{e}e}
-
X_{\nu_{\mu}e}
}
\left[
{\displaystyle
N^{\mathrm{ES}}
\over\displaystyle
\Phi
}
-
X_{\nu_{\mu}e}
\right]
\; .
\label{E573}
\end{equation}
The right-hand side of this relation
contains only measurable and known quantities
(the total flux
$ \Phi $
is given by
Eqs.(\ref{E567}), (\ref{E568}) and (\ref{E569})).
If the right-hand side of Eq.(\ref{E573})
will be found to have a value less than one,
it will be a model independent
proof that solar $\nu_e$'s
transform into other active states.

\begin{figure}[t]
\begin{minipage}[t]{0.49\linewidth}
\begin{center}
\mbox{\epsfig{file=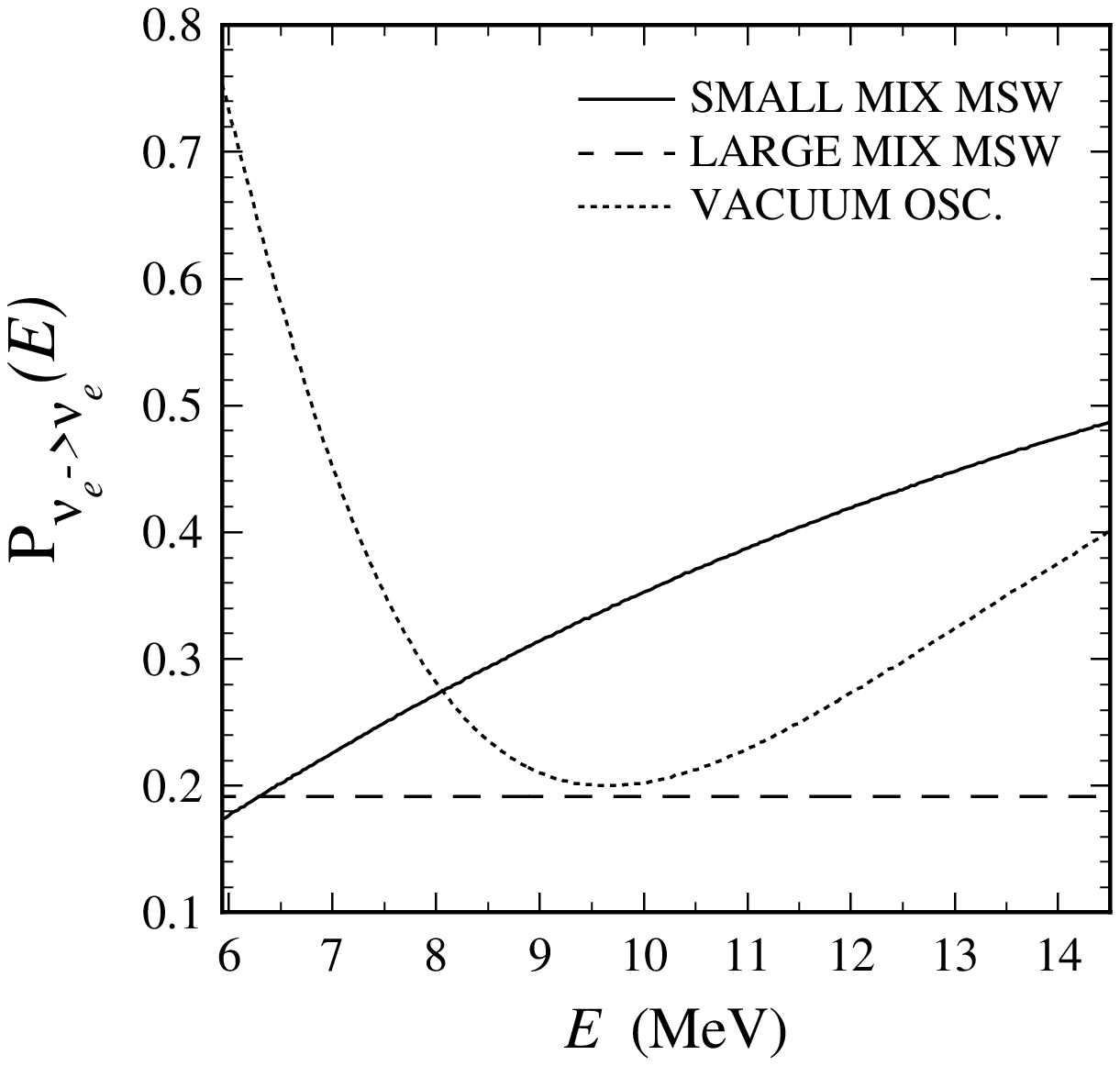,width=\linewidth}}
\end{center}
\protect\caption{\small
Results of the calculations
of the probability of $\nu_e$'s to survive,
$ \mathrm{P}_{\nu_{e}\to\nu_{e}}(E) $,
as a function of the neutrino energy $E$,
in the model with
$\nu_{e}$--$\nu_{\mu(\tau)}$
mixing.
The curves correspond to
small and large mixing angle MSW transitions
and
to vacuum oscillations.
The values of the parameters
$\Delta m^2$ and $\sin^2 2\theta$
used in the calculation
are given in Table~\ref{T:MSW}.}
\label{F10}
\end{minipage}
\hfill
\begin{minipage}[t]{0.49\linewidth}
\begin{center}
\mbox{\epsfig{file=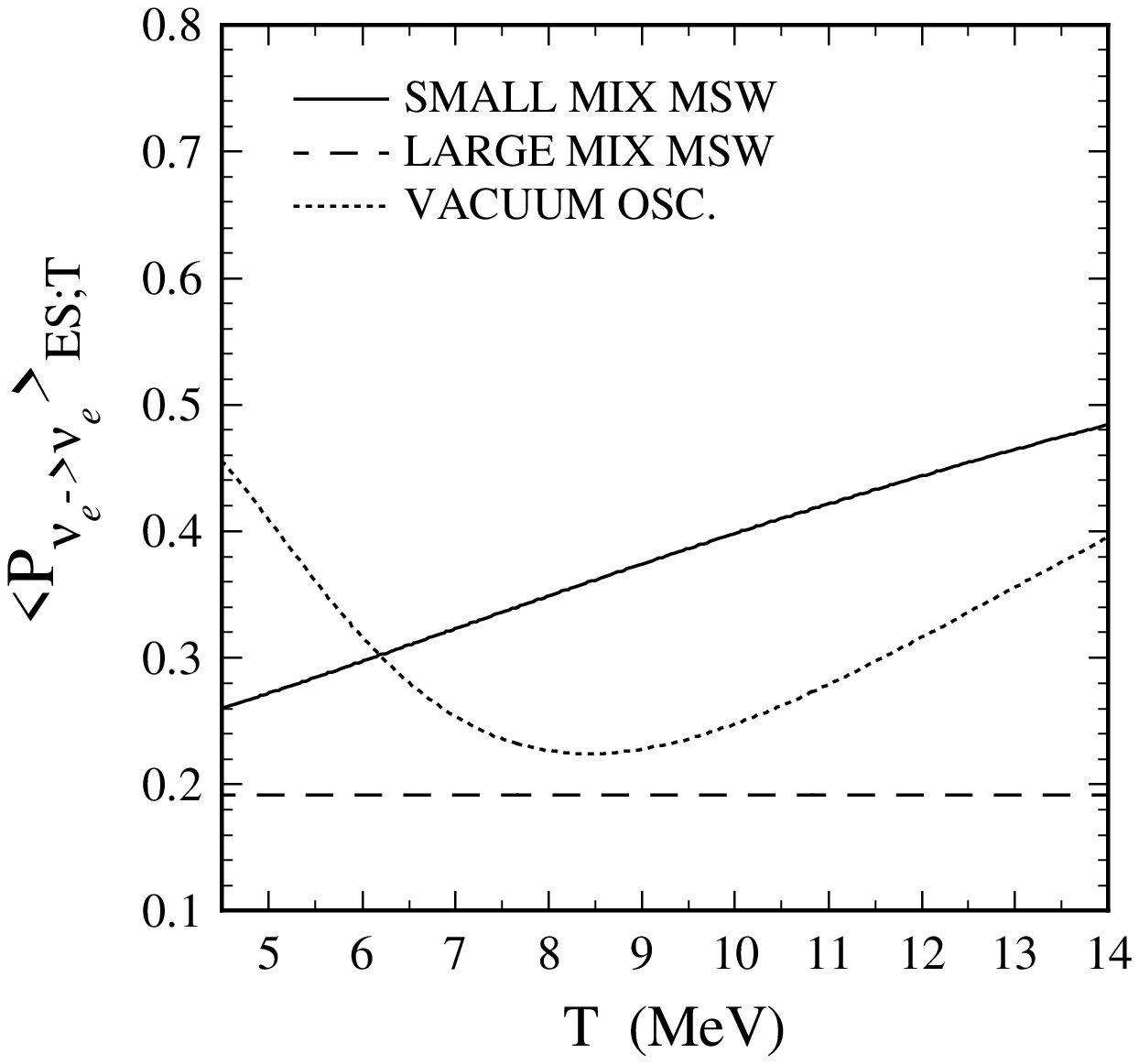,width=\linewidth}}
\end{center}
\protect\caption{\small
Results of the calculation of
$ \displaystyle
\left\langle
\mathrm{P}_{\nu_{e}\to\nu_{e}}
\right\rangle_{\mathrm{ES};\mathrm{T}}
$
(see Eq.(\protect\ref{E574}))
in the model with
$\nu_{e}$--$\nu_{\mu(\tau)}$
mixing
($\mathrm{T}$
is the kinetic energy of the recoil electrons in the ES process).
The curves correspond to
small and large mixing angle MSW transitions
and
to vacuum oscillations.
The values of the parameters
$\Delta m^2$ and $\sin^2 2\theta$
used in the calculation
are given in Table~\ref{T:MSW}.}
\label{F11}
\end{minipage}
\end{figure}

Finally,
a measurement
of the recoil electron spectrum
$ n^{\mathrm{ES}}(\mathrm{T}) $
will allow to determine the average
\begin{equation}
\left\langle
\mathrm{P}_{\nu_{e}\to\nu_{e}}
\right\rangle_{\mathrm{ES};\mathrm{T}}
\equiv
{\displaystyle
1
\over\displaystyle
X_{\nu_{e}e}(\mathrm{T})
-
X_{\nu_{\mu}e}(\mathrm{T})
}
\int_{E_{\mathrm{m}}(\mathrm{T})}
\left[
{\displaystyle
\mathrm{d} \sigma_{\nu_{e}e}
\over\displaystyle
\mathrm{d} \mathrm{T}
}
(E,\mathrm{T})
-
{\displaystyle
\mathrm{d} \sigma_{\nu_{\mu}e}
\over\displaystyle
\mathrm{d} \mathrm{T}
}
(E,\mathrm{T})
\right]
X(E)
\mathrm{P}_{\nu_{e}\to\nu_{e}}(E) \,
\mathrm{d} E
\; ,
\label{E574}
\end{equation}
where
\begin{equation}
X_{\nu_{e}e}(\mathrm{T})
\equiv
\int_{E_{\mathrm{m}}(\mathrm{T})}
{\displaystyle
\mathrm{d} \sigma_{\nu_{e}e}
\over\displaystyle
\mathrm{d} \mathrm{T}
}
(E,\mathrm{T})
\,
X(E)
\,
\mathrm{d} E
\label{E575}
\end{equation}
and
$ X_{\nu_{\mu}e}(\mathrm{T}) $
is given by Eq.(\ref{E533}).
The functions
$ X_{\nu_{e}e}(\mathrm{T}) $
and
$ X_{\nu_{\mu}e}(\mathrm{T}) $
are plotted in Fig.\ref{F04}.
{}From Eq.(\ref{E531})
we easily obtain
\begin{equation}
\left\langle
\mathrm{P}_{\nu_{e}\to\nu_{e}}
\right\rangle_{\mathrm{ES};\mathrm{T}}
=
{\displaystyle
1
\over\displaystyle
X_{\nu_{e}e}(\mathrm{T})
-
X_{\nu_{\mu}e}(\mathrm{T})
}
\left[
{\displaystyle
n^{\mathrm{ES}}(\mathrm{T})
\over\displaystyle
\Phi
}
-
X_{\nu_{\mu}e}(\mathrm{T})
\right]
\; .
\label{E576}
\end{equation}
If the right-hand side of Eq.(\ref{E576})
will be found to have a dependence on $ \mathrm{T} $,
it will be a model independent proof
that there are transitions of solar $\nu_e$'s
into other active neutrinos.

We have calculated the survival probability
$ \displaystyle
\mathrm{P}_{\nu_{e}\to\nu_{e}}(E)
$
and the quantities
$ \displaystyle
\left\langle
\mathrm{P}_{\nu_{e}\to\nu_{e}}
\right\rangle_{\mathrm{ES}}
$
and
$ \displaystyle
\left\langle
\mathrm{P}_{\nu_{e}\to\nu_{e}}
\right\rangle_{\mathrm{ES};\mathrm{T}}
$
in the model with
$\nu_{e}$--$\nu_{\mu(\tau)}$
mixing.
The values of the parameters
$ \Delta m^2 $ and $ \sin^2 2 \theta $
and the values of the quantity
$ \displaystyle
\left\langle
\mathrm{P}_{\nu_{e}\to\nu_{e}}
\right\rangle_{\mathrm{ES}}
$
are given in Table~\ref{T:MSW}.
The values of the mixing parameters
were obtained from a fit of the existing solar neutrino data
under the assumption that
the neutrino fluxes are given by the SSM.
The results of the calculations
of
$ \displaystyle
\mathrm{P}_{\nu_{e}\to\nu_{e}}(E)
$
and
$ \displaystyle
\left\langle
\mathrm{P}_{\nu_{e}\to\nu_{e}}
\right\rangle_{\mathrm{ES};\mathrm{T}}
$
are presented in Figs.\ref{F10} and \ref{F11}.
These figures illustrate
the possibility to distinguish
the different mechanisms of neutrino transitions
through the investigation
of the energy dependence
of the probability of $\nu_e$'s
to survive,
$ \displaystyle
\mathrm{P}_{\nu_{e}\to\nu_{e}}(E)
$,
and of the averaged probability
$ \displaystyle
\left\langle
\mathrm{P}_{\nu_{e}\to\nu_{e}}
\right\rangle_{\mathrm{ES};\mathrm{T}}
$.

\section{Conclusions}

We have shown that the existing solar neutrino data
allow to exclude
in a model independent way
large regions
of the values of the parameters
$ \Delta m^2 $ and $ \sin^2 2 \theta $.
We have also shown that
from future solar neutrino experiments
(SNO, Super-Kamiokande, Icarus)
it will be possible:
1) To reveal in a solar model independent way
the presence of sterile neutrinos
in the flux of solar neutrinos on the earth;
2) To determine directly from the experimental data
the flux of initial $^8\mathrm{B}$ $\nu_e$'s;
3) To obtain directly from the data
the probability of $\nu_e$ to survive.
Thus,
the future solar neutrino experiments
will allow to check the predictions
of the SSM independently
from the properties of neutrinos
and will allow to investigate
neutrino mixing in a model independent way.

\end{document}